\documentclass[12pt , titlepage]{article}

\usepackage{natbib}
\usepackage{amsmath,amssymb,amstext,amsthm,graphics, layout, lscape, longtable}
\usepackage[normalem]{ulem}
\usepackage{color}

\usepackage{url}
\usepackage{caption}
\usepackage{amssymb,amsmath,graphicx,multicol,epsfig,pictexwd,spverbatim}
\usepackage{color, colortbl}
\usepackage{multirow,multicol,dcolumn,booktabs}
\definecolor{mygray}{gray}{0.8}
\usepackage{lscape}
\usepackage[utf8]{inputenc}
\usepackage{graphicx}
\usepackage{hyperref}

\usepackage{tablefootnote}
\usepackage{tabu}
\usepackage{authblk}
\usepackage{array}
\usepackage{subfig}
\usepackage{caption}
\usepackage{bm}
\usepackage{xcolor,colortbl}
\usepackage{multirow}
\usepackage{booktabs}
\usepackage{tabularx}
\usepackage{fourier} 
\usepackage{array}
\usepackage{makecell}
\usepackage{array}
\usepackage{xcolor}



\newcommand{\btheta}{\text{\boldmath{$\theta$}}}
\newcommand{\bkappa}{\text{\boldmath{$\kappa$}}}

\newcommand{\by}{\text{\boldmath{$y$}}}

\newcommand{\bi}{\begin{itemize}}
\newcommand{\ei}{\end{itemize}}

\newcolumntype{C}{@{\extracolsep{.4em}}c@{\extracolsep{0pt}}}%

\RequirePackage[top=1in, left=1in, right=1in]{geometry}

\begin{document}
\providecommand{\keywords}[1]{\textbf{\textit{Key words: }} #1}

\title{Bayesian Meta-Analysis of Penetrance for Cancer Risk}

\author[1]{Thanthirige Lakshika M. Ruberu}
\author[2,3]{Danielle Braun}
\author[3,2]{Giovanni Parmigiani}
\author[1*]{Swati Biswas}

\affil[1]{Department of Mathematical Sciences, University of Texas at Dallas~~~~~~~~~}
\affil[2]{Department of Biostatistics, Harvard T.H. Chan School of Public Health}
\affil[3]{Department of Data Science, Dana Farber Cancer Institute~~~~~~~~~~~~~~~~~~~~~~~~~}
\affil[*]{Correspondence to Swati Biswas, 800 W Campbell Rd, FO 35, Richardson, TX 75025. Email: swati.biswas@utdallas.edu ~~~~~~~~~~~~~~~~~~~~~~~~~}
\date{}

\maketitle

\begin{abstract}
Multi-gene panel testing allows many cancer susceptibility genes to be tested quickly at a lower cost making such testing accessible to a broader population. Thus, more patients carrying pathogenic germline mutations in various cancer-susceptibility genes are being identified. This creates a great opportunity, as well as an urgent need, to counsel these patients about appropriate risk reducing management strategies. Counseling hinges on accurate estimates of age-specific risks of developing various cancers associated with mutations in a specific gene, i.e., penetrance estimation. We propose a meta-analysis approach based on a Bayesian hierarchical random-effects model to obtain penetrance estimates by integrating studies reporting different types of risk measures (e.g., penetrance, relative risk, odds ratio) while accounting for the associated uncertainties. After estimating posterior distributions of the parameters via a Markov chain Monte Carlo algorithm, we estimate penetrance and credible intervals. We investigate the proposed method and compare with an existing approach via simulations based on studies reporting risks for two moderate-risk breast cancer susceptibility genes, ATM and PALB2. Our proposed method is far superior in terms of coverage probability of credible intervals and mean square error of estimates. Finally, we apply our method to estimate the penetrance of breast cancer among carriers of pathogenic mutations in the ATM gene.\\
\\
\end{abstract}

\section{Introduction}
\label{s:intro}

Until recently, genetic testing for cancer-susceptibility genes was limited to a handful of well-known and well-studied genes such as BRCA1 and BRCA2 for hereditary breast and ovarian cancers. However, the advent of next generation DNA sequencing now allows testing a larger number (25 to 125) of cancer genes
at a fraction of the cost and time compared to what was once needed. A landmark decision of the US Supreme Court in 2013 over-turning the patent for genetic testing has further fueled the availability of multi-gene panel testing to a broader population at a cheaper cost \citep{Plichta2016}. These developments, in turn, have set the stage for a paradigm shift in hereditary cancer risk assessment.

As increasing numbers of individuals are being tested via multi-gene panels, more patients are being
identified with pathogenic mutations not only in well-studied genes such as BRCA1 or BRCA2 but also in relatively less-studied genes such as ATM, CHEK2, and PALB2. These patients need help in interpreting their test
results and assessing their risks for various cancers. Indeed there is an unprecedented opportunity to help individuals with pathogenic variants in cancer susceptibility genes whose mutation status was unknown due to lack of testing and therefore were not helped in mitigating their cancer risk. Appropriate management strategies could include, for example, targeted surveillance or prophylactic
options such as mastectomy or oophorectomy, and involve complex decisions for patients and their doctors. This has created an urgent need to streamline the process of counseling
patients especially under the practical constraints of time and resources faced by clinicians. If this target population is appropriately counseled, morbidity and mortality rates associated with germline genetic mutations may be substantially reduced. To effectively counsel patients, a critical tool is
accurate estimate of penetrance, i.e., age-specific risks of developing cancers associated with pathogenic variants in a specific gene.

One of the sources that provides age-specific penetrance is an online risk calculator called ASK2ME (All Syndromes Known to Man Evaluator) \citep{ASK2ME}. ASK2ME takes as input a gene name in which a patient has a pathogenic variant and other relevant information (e.g., age, gender)
and returns age-specific risks for cancers associated with that pathogenic variant. The risk assessment is based on an in-depth literature review for each gene-cancer association. More specifically, for each gene-cancer association, all relevant studies are curated. 
The study deemed to be of highest quality is used for deriving the age-specific risk of developing a specific cancer (i.e., penetrance). Ideally such a study should be a high quality meta-analysis of all relevant studies on that specific gene-cancer association. Yet, such meta-analyses for many gene-cancer associations are unavailable.

The PanelPRO R package is another tool for assessing cancer risk \citep{10.7554/eLife.68699}. The package models comprehensive cancer risk profiles for individuals with family history by integrating a large number of gene mutations and a wide range of potential cancer types to provide age-specific estimates of cancer risk. This calculation requires information about penetrances for the mutations, which are derived from peer-reviewed studies. Again, using meta-analysis estimates of penetrance is ideal for the PanelPRO model.

For many moderate- and high-risk cancer susceptibility genes, there is a substantial literature on the cancer risks that they confer. However, these published studies
typically vary by their designs (e.g., family-based, case-control) and the corresponding summaries/results that they report for risk (e.g., age-specific penetrance, odds ratio (OR), relative risk (RR), standardized incidence ratio (SIR)). This vast amount of scientific knowledge needs
to be synthesized and translated into accurate cancer risk assessment, but this is not trivial.  

In particular, the motivation for this work comes from the need to accurately quantify the penetrance of breast cancer (BC) among carriers of pathogenic variants in the ATM gene, through the synthesis of all relevant evidence in a systematic and coherent manner while allowing for heterogeneity in study populations, study designs, and type of reported results. ATM is a moderate-risk gene that is relatively less-studied compared to the high-risk genes BRCA1/2. Approximately 1-2\% of individuals in the U.S. are believed to carry a risk-conferring ATM mutation \citep{jerzak2018ataxia} and thus its penetrance is critical for clinical management of patients carrying these mutations. There are several published studies on BC risk conferred by ATM and the knowledge base for this gene-cancer association is rapidly expanding \citep{Acevedo2018, Robson2018, Marabelli2016}.

To accomplish the challenging task of synthesizing relevant studies, the most commonly used meta-analysis methods based on either fixed or random-effects models, cannot be used because they require all studies to report the same type of summaries \citep{RandomMeta}. \citet{Marabelli2016} developed a likelihood-based meta-analysis method that can incorporate different types of summaries, and used it for a meta-analysis to estimate BC penetrance for ATM mutation carriers. This estimate is currently used in ASK2ME. This approach is a major step in the direction of synthesizing heterogeneous studies and serves as a stepping stone for our proposed methodology. Nonetheless, it is a fixed-effects approach wherein studies are assumed to be independent and identically distributed (iid) given the penetrance parameters shared by them. It relies on strong modeling assumptions and ignores several sources of uncertainty resulting in potential underestimation of standard error (SE). Further, the approach has not been extensively tested through simulation studies.

Additionally, in their ATM-BC meta-analysis, \citet{Marabelli2016} did not exclude studies that failed to adjust for ascertainment criteria. Studies on cancer risk often tend to ascertain their sample based on family history or age of onset \citep{daly2017nccn} and if that ascertainment information is not appropriately accounted for while estimating the cancer risk by the study (e.g., by conditioning the likelihood on ascertainment factors), the estimate can be  biased \citep{kraft2000bias}. For these reasons, we think it is timely to conduct a meta-analysis to estimate ATM-BC penetrance based on a more refined statistical model and a more rigorous selection of studies including only those studies that appropriately accounted for ascertainment. More specifically, we discard studies without proper ascertainment adjustment and include additional eligible studies published since the work of \citet{Marabelli2016}.

There is a vast amount of literature on Bayesian methods for meta-analysis. \citet{de2015adverse} used a Bayesian zero-inflated random-effect model to estimate adverse effects
associated with a technique for performing total hip arthroplasty. Another study proposed
CPBayes (Cross-Phenotype Bayes) based on a spike and slab prior to simultaneously detect
loci associated with multiple traits (overall pleiotropy) and an optimal subset of associated
traits underlying a pleiotropic signal using GWAS (genome-wide association studies) sum-
mary statistics \citep{majumdar2018efficient}. Bayesian network meta-analysis is another popular
technique in randomized controlled trials to establish the most effective intervention \citep{gao2021new,jepsen2020regenerative}. Nonetheless, there is no Bayesian meta-analysis approach
available that can combine estimates of different types of risk measures.

In fact, there is no \emph{random-effects} meta-analysis approach (in classical or Bayesian framework) that can integrate different types of risk measures
such as age-specific penetrances, OR, RR, and SIR as well as account for uncertainties involved in
such synthesis. Both aspects are important for an accurate synthesis of evidence from a wide range of study types while ensuring that the estimates are robust and valid across a spectrum of clinical scenarios in which risk assessments are typically made in practice. Our proposed Bayesian method will be the first of its kind. The Bayesian framework allows accounting for different sources of uncertainties and relaxing strong modeling assumptions in a natural way via hierarchical modeling and appropriate choice of priors \citep{ParmigianiBook}. Further, studies with different designs and types of results can all be integrated in a seamless manner under one unified model \citep{NetworkBook}. Such modeling allows studies to borrow information from each other resulting in penetrance estimates which are more likely to be accurate and robust. We evaluate the performance of our method and compare it with the method of \citet{Marabelli2016} through simulations  that mimic real studies on ATM-BC association. To further assess the robustness of our method for other moderate-risk genes such as PALB2, we also performed simulations based on PALB2-BC penetrance. Finally, we apply our method to estimate the age-specific penetrance of BC for ATM.

\section{Methods}
\label{s:methods}

\subsection{Likelihood Formulation}
\label{sub:likelihood}
Consider $S$ studies reporting point estimates as well as standard errors in one of four modalities: penetrance (age-specific risk) at some specific ages;
RR; SIR or OR. We assume studies have no patient overlap and are conditionally independent given the study-specific parameters specified below. We use $s$ to index studies. Without loss of generality, we order studies so that the first $S_1$ report penetrances, the next $S_2$ report RR, and so forth.

We assume that, irrespective of the reporting modality, the cumulative penetrance $F_s(t|\kappa_s,\lambda_s)$ at age $t$ for study $s$ is given by the c.d.f. of a Weibull distribution with shape parameter $\kappa_s$ and scale parameter $\lambda_s$. The corresponding density function is $f_s(t|\kappa_s,\lambda_s)$. Jointly, the parameters are denoted by $\btheta_s = (\kappa_s, \lambda_s)$. The Weibull distribution is very popular in survival and reliability analysis as it is able to accurately model time to failure of real life events  in a wide variety of populations \citep{plana2022cancer}, despite having only two parameters.

For each modality, we express the corresponding likelihood $L^{\mbox{P}},L^{\mbox{RR}},L^{\mbox{SIR}},$ and $L^{\mbox{OR}}$ of the reported results in terms of penetrance parameters as in \citet{Marabelli2016}. 
Letting $\btheta = (\btheta_1, \ldots, \btheta_S)$, we write the overall likelihood combining all the studies as

\begin{equation*}
L(\btheta)  = \prod_{s=1}^{S_1} L^{\mbox{P}} (\btheta_s) \prod_{S_1+1}^{\sum_{i=1}^2 S_i} L^{\mbox{RR}} (\btheta_s) \prod_{\sum_{i=1}^2 S_i+1}^{\sum_{i=1}^3 S_i} L^{\mbox{SIR}} (\btheta_s) \prod_{\sum_{i=1}^3 S_i+1}^{\sum_{i=1}^4 S_i} L^{\mbox{OR}} (\btheta_s).
\end{equation*}
We now describe the formulation of the modality-specific likelihood terms.

\subsubsection{Age-Specific Risk (Penetrance) Estimates}
\label{sub:penet}
In this type of study, reported results include a vector $y_s^{\mbox{P}} = ( y_{1s}^{\mbox{P}}, \ldots, y_{ms}^{\mbox{P}} )$ of penetrance values at $m$ ages $(a_1, \ldots, a_m)$, ideally along with a corresponding $m \times m$ covariance matrix $W$ quantifying estimation error. If penetrances are reported by age interval (e.g., decade) $(a_1, \ldots, a_m)$ can be the midpoints. Note that $m$ can vary between studies (see Web Appendix A).

We specify the likelihood by modeling the joint distribution of $\{ \mbox{logit} (y_{1s}^{\mbox{P}}), \ldots, \mbox{logit} 
(y_{ms}^{\mbox{P}}) \}$ given $\btheta_s$ and $W$. A more comprehensive likelihood specification would also acknowledge sampling variability in $W$. This is not considered here, but would be a valuable extension. 
Specifically, we posit the following multivariate normal (MVN) distribution: 
\begin{equation*} 
\bigl\{ \mbox{logit} (y_{1s}^{\mbox{P}}), \ldots, \mbox{logit} 
(y_{ms}^{\mbox{P}}) \bigr\}
\sim \mbox{MVN} \Bigl( \bigl\{ \mbox{logit} ( F_s(a_1 | \kappa_s, \lambda_s) ),\ldots,
\mbox{logit} ( F_s(a_m | \kappa_s, \lambda_s) ) \bigr\}, W^*
\Bigr) = L^{\mbox{P}} (\btheta_s).
\end{equation*}

If $W$ is reported, $W^*$ can be obtained via, say, an application of the multivariate delta method.
Typically, studies do not report $W$ but rather only 95$\%$ confidence intervals (CIs) of the individual components of $y_s^{\mbox{P}}$. In these cases, we take the logit of the lower and upper limits of each  age-specific CI and estimate variances in the diagonal of $W^*$ considering the interval width and assuming normality. The covariances in the off-diagonal elements of $W^*$ are approximated using the numerical method described in \citet{Marabelli2016}.

\subsubsection{Relative Risk} 
In this type of study, reported results include a scalar RR estimate $y_s^{RR}$ with variance $w_{s}$. 
Depending on the penetrance function, the estimated relative risk can vary considerably with age of study participants. To relate the reported RR to the penetrance curve more reliably, we explicitly consider the age distribution of individuals studied. Based on our experience 
we cannot count on studies to report individual-level data, but it is often possible to obtain means $A_1$ and $A_0$ and subject-to-subject variances $V_1$ and $V_0$ for carriers and non-carriers, respectively.
Following \citet{Marabelli2016}, we assume ages of onset among carriers and non-carriers have densities $q_1(a) = N(A_1, V_1)$ and $q_0(a) = N(A_0, V_0)$. We also need to incorporate information about the distribution $F_0(a)$ with density $f_0(a)$ of the penetrance of breast cancer among non-carriers. This can be estimated using registry data, and can be considered known for our purposes. 

We specify the likelihood by modeling the distribution of $\mbox{log} ( y_{s}^{\mbox{RR}} )$ given $\btheta_s$ and $w_s^*$, where $w_s^*$ is an appropriate transformation of $w_s$. We use the log transformation to facilitate the fit of the normal approximation. 
Expressing the mean of the RR approximately in terms of the penetrance, we posit the following distribution:  
\begin{equation}\label{RR}
  \log y_s^{RR} \sim N \Biggl(
  \log \left ( \frac{\int{f_{s}(a|\kappa_s, \lambda_s) q_1(a)da}}{\int{f_{0}(a) q_0(a)da}}
 \right ) ,
  w_s^* \Biggr ).
\end{equation}

The derivations of equation \eqref{RR} as well as equations~\eqref{SIR} and \eqref{OR} (to follow in later sub-sections) can be found in Web Appendix B. As studies typically do not report $w_s$, we obtain $w_s^*$ directly by taking log of the lower and upper limits of the reported CI for RR, computing the width and back-solving using a normal approximation. Note that $q_1(a)$ is unrelated to the assumed Weibull penetrance at age $a$. The latter is the probability of getting cancer by age $a$, i.e., it is conditional on age $a$ while $q_1(a)$ is the marginal distribution of age of onset (of cancer) for carriers (see two equations preceding equation (1) of Web Appendix B).
\subsubsection{Standardized Incidence Ratio}
\label{s:SIR}
In this type of study, reported results include a scalar 
SIR estimate $y_s^{SIR}$ with variance $w_s$. Similar to the RR case, using the log transformation, we express mean of SIR in terms of penetrance approximately using the following distribution. 
\begin{equation}\label{SIR}
  \log y_s^{SIR} \sim N \Biggl(
  \log \left ( \frac{\int{f_{s}(a|\kappa_s, \lambda_s) q_1(a)da}}{P(g=0)\int{f_{0}(a) q_0(a)da}+P(g=1)\int{f_{s}(a|\kappa_s, \lambda_s) q_1(a)da}}
 \right ) ,
  w_s^* \Biggr ),
  \end{equation}
where $P(g=1)$ and $P(g=0)$ are the prevalence of carriers and non-carriers, respectively and $w_s^*$ is obtained similar to as in the RR case. Note that for a rare gene mutation, we may assume that the incidence of BC in the general population is the same as in non-carriers, in which case equation~\eqref{SIR} reduces approximately to equation~\eqref{RR}.

\subsubsection{Odds Ratio}

In this type of study, reported results include a scalar 
OR estimate $y_s^{OR}$ with variance $w_s$. We assume 
\[ \log y_s^{OR} \sim N \bigg(\log \nu_s, \; w_s^*\bigg), \] where $w_s^*$ is obtained similarly to the RR case and $\nu_s$ is an approximation of OR in terms of penetrance given by 
\begin{equation}\label{OR}
\nu_s = \left. \frac{\int{f_{s}(a|\kappa_s, \lambda_s) q_{c1}(a)da}}{\int{f_{0}(a) q_{c0}(a)da}} \middle/\frac{\int{(1-F_{s}(a|\kappa_s, \lambda_s)) q_{h1}(a)da}}{\int{(1-F_{0}(a))q_{h0}(a)da}} \right.,
\end{equation}
where $q_{c1}$ and $q_{c0}$ are distributions of ages of onset for cases while $q_{h1}$ and $q_{h0}$ are the corresponding distributions of ages at inclusion in the study for healthy controls. We assume $q_{c1} = N(A_{c1}, V_{c1})$, $q_{c0} = N(A_{c0}, V_{c0})$, $q_{h1} = N(A_{h1}, V_{h1})$, and $q_{h0} = N(A_{h0}, V_{h0})$; here, 1 and 0 in the subscript indicate carriers and non-carriers, respectively.


\subsection{Prior Distributions}
\label{sub:prior}
We assume that our study-specific $\kappa_s$ and $\lambda_s$ parameters follow Gamma priors and the hyper-parameters follow uniform distributions. In particular, we consider the following hierarchical priors: $\pi(\kappa_s|a, b) =$ Gamma$(a, b), \;  \pi(\lambda_s|c, d) =$ Gamma$(c, d),\;$ where $a$ and $c$ are shape parameters, $b$ and $d$ are scale parameters, and $\pi(a|l_a, u_a) =$ U$(7.5,27.5), \; \pi(b|l_b, u_b) =$ U$(0.15,0.25), \;  \pi(c|l_c, u_c) =$ U$(43,63), \;$and$ \; \pi(d|l_d, u_d) =$ U$(1.32,2.02)$. A detailed description of the process used to determine the limits of the uniform distributions can be found in Web Appendix C.

\subsection{MCMC Algorithm}
\label{sub:MCMC}
We implement a Markov chain Monte Carlo (MCMC) algorithm for estimating posterior distributions. For this, we employ a standard (non-adaptive) Metropolis-Hastings algorithm within Gibbs sampling. As is the case with our prior specification, the algorithm is applicable to most gene-cancer combinations. We run 30,000 MCMC iterations with 15,000 burn-in. We carry out all analysis in statistical software system R \citep{R}. Algorithm details and convergence statistics are provided in Web Appendix D.

\noindent
{\bf {Consensus Penetrance:}} Recall that the method assumes that irrespective of the modality, the reported risk estimate follows a normal distribution whose mean is expressed in terms of Weibull distribution (and hence its parameters $\kappa_s$ and $\lambda_s$) and variance is fixed at what was reported by the study. The $\kappa_s$ and $\lambda_s$ parameters are then assumed to follow gamma priors with their parameters having uniform hyper-priors. After estimating the posterior distributions of all parameters, we estimate the final meta-analysis penetrance curve. Specifically, at iteration $t$: (1) We compute $\kappa^{(t)}=a^{(t)}*b^{(t)}$ and $\lambda^{(t)}=c^{(t)}*d^{(t)}$ and (2) Use the Weibull($\kappa^{(t)},\lambda^{(t)}$) cdf at ages 40, 50, 60, 70, and 80 as penetrance estimates at the $t^{th}$ iteration. Finally, for each age, the mean of the penetrance values over all iterations are computed, which serves as the meta-analysis estimate. We also obtain credible intervals (CrI) using these posterior distributions of penetrances at each age. 

For ease of readability, we have listed the assumptions and details of the proposed method (as well as simulations and application to follow) in Web Appendix A.

\section{Simulation Study}
\label{s:Sim}


We investigate the performance of the proposed method through simulations. Further, we compare our method with the  approach of \citet{Marabelli2016} (see Web Appendix E for details of this method). To evaluate the performance of our method under different BC susceptibility genes, we consider two settings for simulations, one based on the ATM gene and another based on the PALB2 gene. In combination, these two sets of simulations will illustrate the broad scope of our method.

As described in Section ~\ref{s:SIR}, for studies reporting SIR, we need the prevalence of the pathogenic gene mutations for equation ~\ref{SIR}. However, for many moderate BC penetrance genes, prevalence has not been well established. Therefore, following \citet{Marabelli2016}, for studies providing SIR, we use equation~\eqref{RR} (used for studies reporting RR) under the assumption that incidence of BC in the general population is the same as in non-carriers, which is reasonable for rare gene mutations.

We require several distributions of age, namely, $q_1$ and $q_0$, for  equation ~\eqref{RR} and $q_{c1}$, $q_{c0}$, $q_{h1}$, and $q_{h0}$ for equation ~\eqref{OR} as described in Section ~\ref{sub:likelihood}. These are assumed to be study-specific normal distributions with pre-specified mean and variance. In our actual meta-analysis (to follow in Section ~\ref{s:app}), for a given study, we rely on the mean and variance of these distributions reported in that study. However, not all RR and OR studies report  age-related statistics. For example, Table ~\ref{meta} lists 17 studies that we use in our application among which five report RR. Out of these five, four studies report mean and SD of age of onset for carriers ($q_1$); of these four, only two report mean and SD of age of onset for non-carriers ($q_0$). The fifth study did not report any age-related statistics. Similarly, some studies reporting OR do not report any relevant age-related statistics. 

To mimic such real scenarios in our simulations, we assume some studies report age-related statistics but not all. Therefore, although our data generation process (to follow in Section ~\ref{sub:data generate}) allows obtaining relevant age-related summaries for all simulated studies, before applying the meta-analysis method, we replace some of the generated summaries by a fixed mean and SD. That is, we let some studies in our simulations not report the relevant age-related summaries (to mimic reality) and fill in those with a mean of 63 and an SD of 14.00726. These are the mean and SD of age of onset of BC in the US population obtained from Surveillance, Epidemiology, and End Results (SEER) program \citep{seer2}.  Due to lack of more specific information and given that pathogenic variants in ATM or PALB2 are rare,  we assume this mean age and SD for all studies that do not provide age-related distributions $q_1$, $q_0$, $q_{c1}$, $q_{c0}$, $q_{h1}$, and/or $q_{h0}$.

Moreover, case-control studies usually only provide the mean age at diagnosis for cases and the mean age for controls. In fact, none of the ten OR studies listed in Table ~\ref{meta} report mean age of onset for carrier cases and non-carrier cases separately or separate mean age at inclusion in the study for carrier and non carrier controls. Thus, for case-control studies we set $q_{c1}(a) = q_{c0}(a)=q_{c}(a)$ (age of onset for cases) and $q_{h1}(a) = q_{h0}(a)=g_{h}(a)$ (age at inclusion in the study for healthy controls) before applying the Bayesian and \citet{Marabelli2016} methods.

In addition to the above scenario wherein we mimic the observed pattern in the literature, we also consider a simulation scenario with even less information available. In this scenario we assume none of the studies report any age-related statistics. In other words, for each of the two simulation settings (ATM and PALB2) we consider two scenarios. Scenario 1 mimics a realistic situation  in the literature wherein only some RR and OR studies report the relevant age-related summaries and  Scenario 2 serves as a situation with no age-related information available from any RR and OR studies. In Scenario 1, we use the fixed mean $63$ and SD $14.00726$ for any missing (unreported) age-related summary. Under Scenario 2, all age-related summaries for all studies $q_1$, $q_0$, $q_{c}$, and $g_{h}$ are assumed to have the same fixed mean and SD of $63$ and  $14.00726$, respectively. Note that the above age-related statistics are used as inputs for both Bayesian and \citet{Marabelli2016} methods.

\subsection{Simulations based on ATM}
For the simulations based on ATM-BC associations, we borrow information from the 17 ATM-BC association studies identified and used in our meta-analysis data application (Section ~\ref{s:app}) and listed in Table~\ref{meta}. Specifically, for each simulation replicate, we simulate 17 studies whose sample sizes and reported risk measures (penetrance, RR, or OR) are the same as to those given in Table~\ref{meta}.

\subsubsection{Data Generation Model}
\label{sub:data generate}
Each simulation replicate consists of two studies reporting penetrance, five reporting RR, and ten reporting OR (following Table~\ref{meta}). We generate time to BC for carriers using Weibull$(\kappa_s,\lambda_s)$ and a censoring time from $N(85,10)$. The study-specific $\kappa_s$ and $\lambda_s$ are generated from $N(4.55,0.525)$ and $N(95.25,12.375)$, respectively. The choice of parameters is  based on the observed range of $\kappa$ and $\lambda$ values of approximate Weibull curves fitted to each of the 17 studies; the details are explained in the Web Appendix F. Depending on whether a study reports age-specific penetrance, RR, or OR, we generate data and the corresponding measure in the following manner.
\\
{\bf{\emph{Age-Specific Penetrance}}}$\\$
As mentioned above, we first generate two time points for each carrier in the $s^{th}$ study: (1) Time to BC and (2) Censoring time. The minimum of these two times and age 95 (assumed to be the maximum observed age in a study), and an indicator of censoring are considered to be the observed data for each carrier. Next, we fit a Kaplan-Meier curve and get penetrance estimates and their CIs at ages 40, 50, 60, 70, and 80. Note that we do not generate non-carriers because they are not needed for estimation of penetrance. 
\\
{\bf{\emph{RR and OR}}}$\\$
First, we generate a population of 2 million consisting of carriers and non-carriers with carrier probability of 0.01. Then, cancer status for each carrier is generated in the same manner as described above for studies reporting penetrance. In particular, a subject is affected if her time to BC is equal to the minimum of the three times (time to BC, censoring time, and 95) otherwise she is unaffected. Next, the process is repeated for non-carriers by generating time to BC using a truncated Weibull$(3.65, 143.2426)$ distribution, which is an approximation to BC risk estimates given by SEER. The Weibull distribution is truncated at age 185 to ensure that there are enough non-carrier cases at ages less than 80 such that the resulting RR and OR estimates are not excessively large (Web Figure 1 shows the distributions used in the data generation). Finally, for studies reporting RR, the estimate RR and its standard error are obtained for a randomly chosen sample of carriers and non-carriers. 

When sampling we use the sample sizes
of carriers and non-carriers in the studies listed in Table \ref{meta}. For studies reporting OR, the estimate OR and its standard error are calculated for a randomly chosen sample of cases and controls with sample sizes the same as those of the ten OR studies listed in Table \ref{meta}. Further, the age at which a given subject was either diagnosed, censored, or healthy is known at this point of data generation along with their carrier status. Using these information, we construct relevant normal age distributions for a study that does report age-related distributions under Scenario 1 by taking the mean and variance of the corresponding ages. However, when generating OR studies, our data generation process tends to yield a considerably higher mean and lower SD for controls than those for cases. As in most real studies the cases and controls are matched by age, we therefore set the generated mean and SD for the cases same as the corresponding ones for the controls. In Scenario 1, whenever a study (out of 17 listed in Table \ref{meta}) did not report a given age-related distribution, that distribution was replaced by $N(63,14.00726)$ whereas in Scenario 2, we use $N(63,14.00726)$  for all age-related distributions for all studies.

The estimates and their standard errors obtained from all studies along with the relevant age distributions are then input to the Bayesian and \citet{Marabelli2016} methods to obtain overall meta-analysis estimates of penetrance and their CrIs. We generate 500 replicates for each combination of setting and scenario. Then we report average of the estimates, their mean square error (MSE), and coverage probabilities of $95\%$ CrIs.

To evaluate the results of the simulations from both our proposed approach and the approach by \citet{Marabelli2016}, we need to know the true penetrance values. Note that as each study has its own penetrance curve whose study-specific parameters $\kappa_s$ and $\lambda_s$ were generated from common normal distributions, the true (overall) penetrance values (to which the simulation estimates need to be compared to) are not directly
available. Thus, to obtain the true values, we generate a large number of $\kappa_s$ and $\lambda_s$ from their respective normal distributions (which were used in data generation process as described above). Then, for each pair of ($\kappa_s$, $\lambda_s$), we obtain a large number of Weibull penetrance values at each age and compute their mean for each age. These mean penetrance values serve as true values for evaluation of the simulation results.

\subsection{Simulations based on PALB2}
\label{sub:simPALB2}
To check that our proposed method is generally applicable to other gene-cancer associations as well, we conduct simulations based on PALB2-BC association. To obtain a realistic range of risk estimates for this association, we utilized four papers --- two reported age-specific penetrance, one reported RR, and another reported OR \citep{PALB2_meta,Antoniou2014, Casadei2011,Erkko2008}. The type and numeric values of the reported estimates in these four studies vary substantially from each other. In particular, the penetrance at age 70 reported by the two penetrance studies are 35$\%$ and 40$\%$ while the other two studies report RR = 3.4 and OR = 21.4. Based on these four studies, the normal distributions we use for data generation are $\kappa_s \sim N(3.7,0.35)$ and $\lambda_s \sim N(84.5,7.25)$ (more details can be found in the Web Appendix B).

As four studies may not be sufficient for a meaningful meta-analysis and conducting an exhaustive literature review for PALB2-BC association is beyond the scope of this work, we let a simulation replicate consist of 12 studies. Among these, we assume four studies report age-specific penetrance ($n$ = 1000, 500, 1000, 500), four studies report RR ($n$ = 1000, 1800, 2500, 5000), and four studies report OR ($n$ = 500, 1000, 3000, 5000). Similar to the procedure described in Section ~\ref{sub:data generate}, we use the above-mentioned normal distributions to generate the 12 study-specific $\kappa_s$ and $\lambda_s$ for each simulation replicate, based on which we further generate age-specific penetrance, OR, and RR in the same manner as described in the same section. Note that the four PALB2 studies described in the previous paragraph were utilized only to derive the parameters of the normal distributions used in the data generation. After deriving the parameters, the four studies are not used further in any manner. 

Under Scenario 1, only the first two studies within each set of four RR and four OR studies are assumed to have reported the relevant age-related summaries, whereas under Scenario 2, we assume none of the RR and OR studies report the relevant age-related summaries.

\subsection{Simulation Results}
\label{sub:simresults}
The results for Scenarios 1 and 2 based on ATM-BC associations are presented in Table~\ref{atm} and Web Table 1, respectively. Compared with the method of \citet{Marabelli2016}, two advantages of the Bayesian method stand out: (1) MSEs are smaller and (2) coverage probabilities are closer to $95\%$. The latter is the most pronounced as the method of \cite{Marabelli2016} suffers from severe under-coverage.  
Table ~\ref{palb2} contains the simulation results for Scenario 1 based on PALB2 gene. The results for Scenario 2 are in Web Table 2. These results are similar to what we found for Setting 1 (ATM-based) except slightly higher coverage of 95$\%$ CrI. We also investigated the sensitivity of our results to the choice of fixed hyper-parameters, different age-related distributions, and varying the values of $m$ (the number of time points at which penetrances are reported by different studies). Details and results of the sensitivity analyses are reported in Web Appendix G.

\begin{table}
 \centering
\caption{Simulation results for Setting 1 (based on ATM) and Scenario 1 }
\label{atm}
\begin{tabular}{lccccc}
  \hline
Age & 40 & 50 & 60 & 70 & 80\\ \hline
True Penetrance& 0.026 &	0.067&	0.141&	0.253&	0.398\\
Estimated Penetrance (Bayesian) & 0.028&	0.068&	0.140&	0.251&	0.401\\
Estimated Penetrance (Marabelli) &  0.042&	0.084	&0.145&	0.225&	0.323\\
MSE (Bayesian) & 0.0002&	0.0006&	0.0013&	0.0026&	0.0043\\
MSE (Marabelli) & 0.0018	&0.0039	&0.0058	&0.0081&	0.0135\\
95\% CrI coverage (Bayesian) &0.986&	0.98&	0.968&	0.94&	0.932\\ 
95\% CrI coverage (Marabelli)& 0.154	&0.146	&0.100&	0.076&	0.054\\\hline
\end{tabular}
\end{table}

\begin{table}
 \centering
\caption{Simulation results for Setting 2 (based on PALB2) and Scenario 1 }
\label{palb2}
\begin{tabular}{lccccc}
\hline
Age & 40 & 50 & 60 & 70 & 80\\ \hline
True Penetrance& 0.066	&0.143&	0.257&	0.404&	0.565\\
Estimated Penetrance (Bayesian) & 0.071	&0.151&	0.272&	0.426&	0.594\\
Estimated Penetrance (Marabelli) & 0.069&	0.153&	0.269&	0.405	&0.547\\
MSE (Bayesian) & 0.0002&	0.0005&	0.0011&	0.0021&	0.0029\\
MSE (Marabelli) &  0.0003& 	0.0010& 	0.0017& 	0.0022& 	0.0028\\
95\% CrI coverage (Bayesian) &1.00&		1.00&		1.00&		0.996&		0.988 \\ 
95\% CrI coverage (Marabelli)&  0.230&	0.178	&0.136	&0.104&	0.170\\\hline
\end{tabular}
\end{table}

\section{Meta-Analysis of ATM-BC Penetrance}
\label{s:app}

Here, we apply the proposed Bayesian method to estimate the age-specific penetrance of female BC among carriers of pathogenic variants in the ATM gene.

\subsection{Selection of Studies }
We started by considering all studies utilized in \citet{Marabelli2016}, which were primarily identified using a Pubmed search of following keywords in the title/abstract of the articles: [‘‘ATM’’] AND [‘‘penetrance’’ OR ‘‘risk’’] AND [‘‘breast’’] up to February 10, 2015. We extended the same search to cover until December 20, 2021. Additional studies were identified by a semi-automated natural language processing–based procedure for abstract screening \citep{deng2019validation}. Our inclusion criteria for studies is similar to those of \citet{Marabelli2016}, which included (1) family-based segregation analyses or epidemiological studies reporting cancer risk information, in terms of age-specific penetrance, RR, or SIR; (2) case-control studies comparing BC patients with healthy subjects and reporting either OR or sufficient data to estimate the OR and its 95$\%$ CI. However, unlike \citet{Marabelli2016}, we exclude case-control studies that ascertained subjects based on family history of BC but did not adjust for this ascertainment criteria in their analysis. This eliminated eight out of 19 studies reporting OR that were included in  \citet{Marabelli2016}. On the other hand, we include five additional studies reporting OR estimates  and one study reporting RR, which were published after  \citet{Marabelli2016}. We only included pathogenic variants for a study, whenever that information was available. Some studies included information on both pathogenic variants and variants of uncertain significance. The latter are variants for which the pathogenity is unknown; for such studies, we only included counts of pathogenic variants when computing the risk estimates. Moreover, pathogenicity of a given mutation  can change over time in light of new findings. Therefore, whenever a study reported sufficient information, we tried to confirm whether a given variant initially classified as pathogenic remained pathogenic by using ClinVar \citep{clinvar}, a freely available NIH archive of reports of human genetic variants. Our meta-analysis includes 17 studies that are summarized in Table ~\ref{meta}.  Following \citet{Marabelli2016}, we made several assumptions specific to meta-analysis of the ATM gene due to the limited information provided by individual studies and to incorporate as many studies as possible in our meta-analysis (See Web Appendix H).
    
\begin{table}
\scriptsize
 \centering
\caption{Summary of studies included in the meta-analysis of ATM-BC penetrance}
\label{meta}
\begin{tabular}{llcccc}
  \hline
 &\makecell{Study} & \makecell{Cases} & \makecell {Study Design} & \makecell{Sample Size} & \makecell{Risk and CI\\ (Input for \\meta-analysis)}\\\hline \hline
 1& \cite{goldgar2011rare} &\makecell{ Familial BC\\ BRCA1/2 negative}&\makecell{Family-based \\segregation analysis}&156 &\makecell{Penetrance\\ Curve \\ (at ages 35 - 80)} \\ 

  2& \cite{thompson2005cancer} & \makecell{One family member \\with ATM}&\makecell{Cancer incidence in \\relatives of ATM patients} &1160&\makecell{Penetrance\\ Curve \\ (at ages 40 - 80)} \\
  
3& \cite{swift2008breast} &\makecell{One family member \\with ATM}& \makecell{Cancer incidence in \\relatives of ATM
patients}  & 919&\makecell{RR = 2.4\\ (1.3 - 4.3) } \\

4& \cite{renwick2006atm} & \makecell{ Familial BC\\ BRCA1/2 negative}& \makecell{Case-control/family-based
\\segregation analysis}  & 5173&\makecell{RR = 2.37\\(1.51 - 3.78)}\\ 

5& \cite{Li2016} & \makecell{Familial BC\\ BRCA1/2 negative} &\makecell{family-based
\\segregation analysis}& 660&\makecell{RR = 2.67\\(0.82 - 10.56)}\\
             
6&\cite{olsen2005breast} & \makecell{One family member \\with ATM} &\makecell{Cancer incidence in \\relatives of ATM
patients}& 712&\makecell{SIR = 2.9 \\ (1.9 - 4.4)} \\ 
             
7& \cite{andrieu2005ataxia} & \makecell{One family member \\with ATM} &\makecell{Cancer incidence in \\relatives of ATM
patients}& 708&\makecell{SIR = 2.43\\(1.32 - 4.09)}\\

8&  \cite{Kurian2017} & \makecell{Female patients who \\underwent panel testing} &\makecell{Case control study}&95561&\makecell{OR = 1.74 \\(1.46 - 2.07)}\vspace{0.1in}\\

9& \cite{momozawa2018germline} & \makecell{Unselected BC } & \makecell{Case-control study}  & 18292&\makecell{OR = 2.10\\(1.0 - 4.1)}\\ 
            
10&\cite{dorling2021breast} &  \makecell{Mainly unselected BC with\\ a subset of early onset BC \\ BRCA1/2 negative} & \makecell{Case-control study} &97997& \makecell{OR = 2.10$^a$\\ (1.71 - 2.57)} \\ 
             
11& \cite{hu2021population}  & \makecell{Unselected BC } & \makecell{Case-control study} & 64791 & \makecell{OR = 1.82$^{b,d}$\\(1.46 - 2.27)}\\ 
            
12& \cite{mangone2015atm}  & \makecell{Sporadic BC} & \makecell{Case-control study} & 200& \makecell{OR = 3.03$^{c,d,f}$\\(NA)} \\ 
             
 13& \cite{brunet2008atm}  & \makecell{Unselected early-onset \\BC ($<46$ years)\\
BRCA1/2 negative} &\makecell{Case-control study} & 193&\makecell{OR = 18.13$^{c,d}$ \\ (NA)}\\
             
14& \cite{pylkas2007evaluation}  & \makecell{Familial and \\ unselected BC }  & \makecell{Case-control study}&2231&\makecell{OR = 6.93$^{e,f,g}$\\ (0.85 - 56.43)}\\ 
            
15& \cite{Zheng2018}  & \makecell{Unselected BC} &\makecell{Case-control study}&2133&\makecell{OR = 4.40$^f$\\(0.51 - 37.75)} \\ 
             
16& \cite{kreiss2000founder}  & \makecell{Unselected BC} &\makecell{Case-control study}&298&\makecell{OR = 3.09$^f$\\(0.50 - 18.96)} \\ 
             
17&  \cite{fitzgerald1997heterozygous}& \makecell{Early-onset BC \\($<40$ years)} &\makecell{Case-control study}&603&\makecell{OR = 0.50$^f$\\(0.07 - 3.58)}\\ \hline
 \multicolumn{6}{l}{\scriptsize $^a$ Based on 30 studies in breast cancer association consortium (BCAC) unselected for family history.} \\ 
 \multicolumn{6}{l}{\scriptsize $^b$ Based on 12 studies in the CARRIERS consortium not enriched with patients with a family history or early onset of disease.}\\
\multicolumn{6}{l}{\scriptsize $^c$ No mutations in controls, $^d$ Excluded VUSs, $^e$ Discarded subset with familial BC when calculating OR. }\\
\multicolumn{6}{l}{\scriptsize $^f$  OR and CI were calculated using data reported in the paper.}\\
\multicolumn{6}{l}{\scriptsize $^g$  Reported number of mutations for both familial and unselected cases. Only unselected cases were included.}\\
\end{tabular}
\end{table}


\subsection{Results}
Figure 1(a) shows the consensus penetrance of BC for individuals with an ATM pathogenic variant obtained from the Bayesian meta-analysis. The risk of BC is $6.8\%$ by age 50  ($3.4\%-11.6\%$) and $29.5\%$ by age 80 ($21.9\%-38.1\%$). For comparison, the figure also shows the penetrance estimates for non-carriers obtained using SEER data. We also apply the \citet{Marabelli2016} method to the same set of 17 studies. The resulting penetrance curve is lower than the one given by the proposed Bayesian method as shown Figure 1(b). The figure also shows the penetrance curve reported in \citet{Marabelli2016} based on their original 19 studies. To gain some insight into how the final Bayesian meta-analysis curve compares with the estimates provided by the individual 17 studies, in Figure ~\ref{weibull}, we separately plot the approximate Weibull penetrance curve for each study listed in Table \ref{meta}. We also superimpose the penetrance curve given by the \citet{Marabelli2016} method applied on the 17 studies. The Bayesian consensus penetrance curve lies more or less at the center of the curves. However, we note that in addition to the risk estimate reported by a study, the sample size and variance of the estimate also dictate how much contribution an individual study makes to the final meta-analysis estimate.

Within our 17 selected studies, three studies that reported OR included early-onset BC cases \citep{fitzgerald1997heterozygous,brunet2008atm,dorling2021breast} while one study included controls with ages greater than 60 \citep{momozawa2018germline}. We investigate how sensitive our results are to these studies by removing them one at a time. The four penetrance curves obtained in this manner are very close to the curve based on all 17 studies (see Web Figure 2).
\vspace{-12pt}
\begin{figure}
    \centering
    \subfloat[]{{\includegraphics[width=0.45\textwidth]{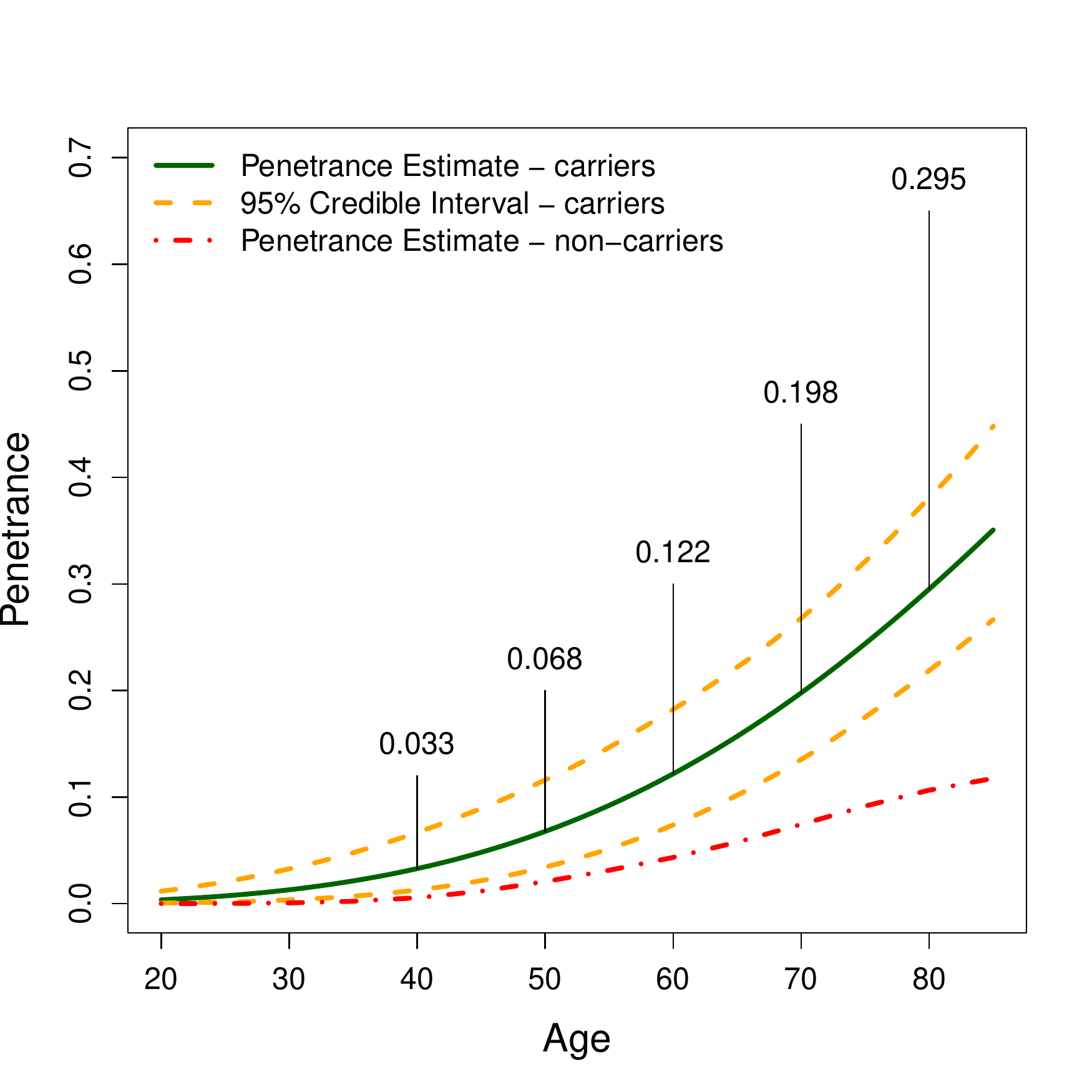} }}
    \qquad
    \subfloat[]{{\includegraphics[width=0.45\textwidth]{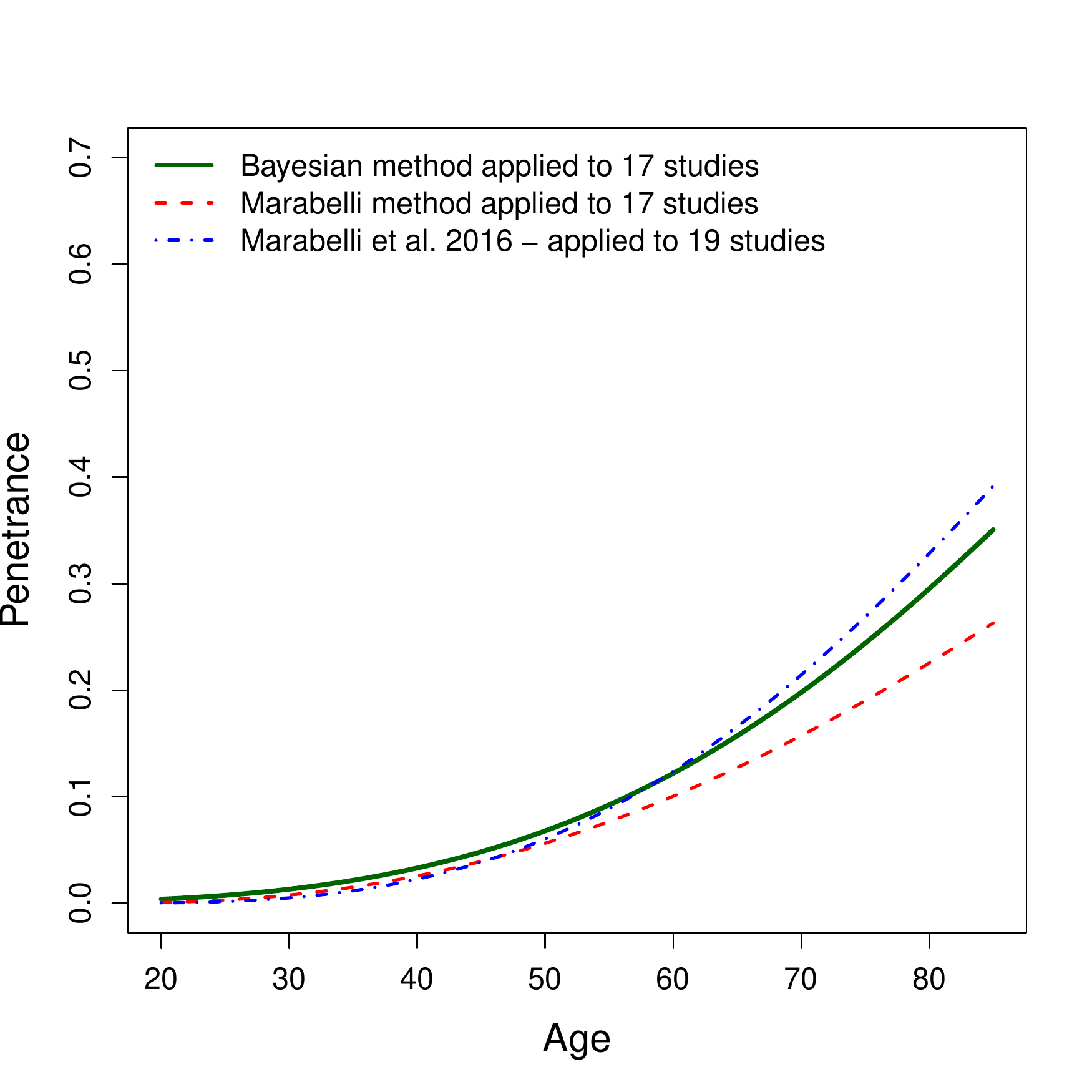} }}
    \caption{(a): Penetrance estimate for ATM-BC. (b): Comparison of the penetrance curves }
    \label{application}
\end{figure}

\begin{figure}[htp]
    \centering
    \includegraphics[width=0.45\textwidth]{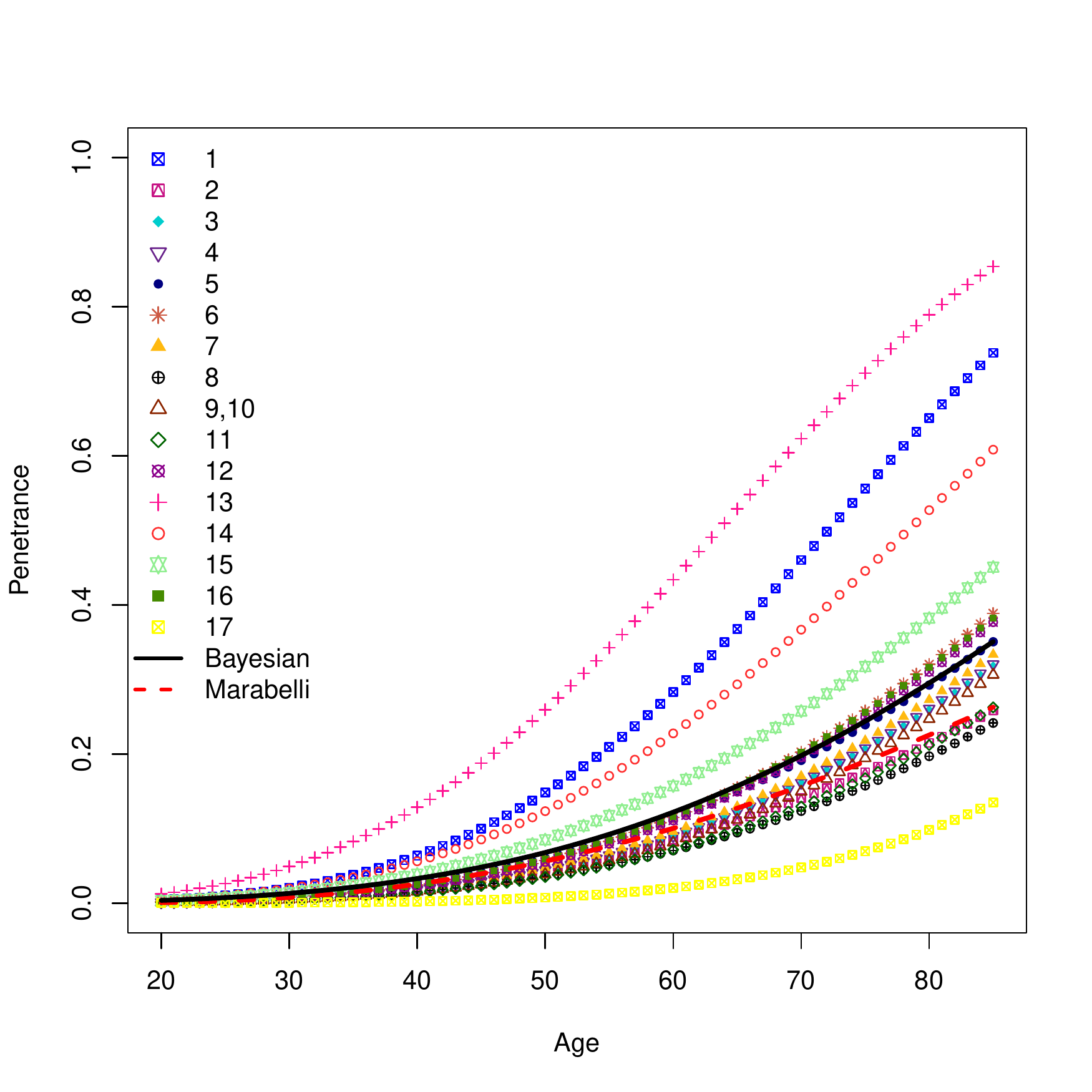}
    \caption{Approximate Weibull distributions for the risk estimates from 17 studies. The number in the legend corresponds to the study index in Table ~\ref{meta}. }
    \label{weibull}
\end{figure}

\section{Discussion}
\label{s:discuss}

The availability of multi-gene panel testing, wherein several cancer-susceptibility genes can be tested using next-generation sequencing, is shifting the paradigm of hereditary cancer risk assessment \citep{Plichta2016}. Panel testing allows identification of patients carrying pathogenic germline mutations in many cancer-susceptibility genes, bringing attention to not only high-risk genes such as BRCA1/2 but also moderate-risk genes like ATM, CHEK2, and PALB2 \citep{turnbull2008genetic}. Indeed, many studies are available that provide cancer risks associated with a particular gene mutation.  However, their study designs and the reported risk measures can vary. Here we propose a novel Bayesian hierarchical random-effects model that integrates the information available from these different types of studies to provide a combined age-specific penetrance for carriers of pathogenic variants while taking into account the uncertainties involved in such an integration. Our simulation study shows that this approach provides estimates close to the true value with smaller MSE than the ones given by the  \citet{Marabelli2016} approach. Moreover, the coverage probabilities of $95\%$ CrI remain around $95\%$.

To the best of our knowledge, \citet{Marabelli2016} is the only study that conducted a meta-analysis by synthesizing studies reporting different types of risk measures. Notwithstanding this ground-breaking  contribution, being a fixed effect meta-analysis approach, it does not allow for heterogeneity across the studies. Indeed, this is reflected in our simulation study wherein 
the coverage probability of CrIs given by this method is extremely low. 

In our ATM-BC meta-analysis, we included only those studies that made an appropriate ascertainment adjustment or studies where such an adjustment was not necessary (population-based studies). Failure to account for study ascertainment may introduce bias in the final estimates \citep{ranola2019exploring}. Even though we did include four studies reporting OR where cases were ascertained by early age of onset and a study with older controls, our sensitivity analysis shows that our estimates are robust to such studies (refer to Web Figure 2). 
It should be noted that none of the RR/OR studies included in our ATM-BC meta-analysis report age-specific estimates. Moreover, in the two studies that reported penetrance curves \citep{goldgar2011rare,thompson2005cancer}, the earliest age at which we could obtain estimates for penetrance was 35. Therefore, caution is warranted in interpreting the reported meta-analysis penetrance curve for ages less than 35. 

As a result of our choice of priors that allow a wide spectrum of penetrance values, the
proposed method can be readily employed for a broad range of gene-cancer combinations.
Indeed, we have applied our method to estimate the age-specific risk for PALB2-BC (manuscript under preparation). Our final meta-analysis penetrance estimates of breast
cancer among carriers of ATM and PALB2 genes have been integrated into a recently
released version 1.0.0 of PanelPRO \citep{10.7554/eLife.68699} and thus can be immediately used for clinical
purposes through the PanelPRO tool. They can be also included in ASK2ME \citep{ASK2ME}. Moreover,
our R package BayesMetaPenetrance (link in Supporting Information) implementing the proposed method can be readily applied to other gene-cancer combinations.

We conclude by acknowledging some limitations of our work. First, our methodology does not directly account for ascertainment: studies that do not report ascertainment-adjusted estimates need to be excluded, unless individual level data are available to perform an adjusted analysis which can then be used as an input. Ascertainment adjustment based on summary information alone is quite challenging: improvements will be possible when more systematic databases of gene-cancer association become available. Second, our methodology does not account for sampling variability in variance estimates reported by the studies. However our simulations suggest that the coverage does not suffer, at least in our scenarios. Our simulations are rooted in real scenarios, but do assume Weibull data-generating mechanisms and normally distributed age distributions throughout. Our empirical analysis of ATM inherits limitations of the reporting practice of the studies considered, particularly with regard to the age distributions of recruited subjects. This limitation will hopefully be mitigated by more extensive reporting in future studies. 

In summary, we provide the first random-effects meta-analysis approach that can integrate different types of risk measures
such as age-specific penetrances, OR, RR, and SIR as well as account for several of the relevant uncertainties involved in
such synthesis. We hope that this work will contribute to more efficient and accurate decision support for carriers of inherited susceptibility to cancer and other diseases.

\section*{Acknowledgements}

This work is supported by NIH grant R03CA242562-01. We are thankful to the associate editor and two referees for their constructive comments and suggestions. We also thank Stephen Knapp for his help in checking pathogenicity status of some ATM mutations. 

\bibliography{arxiv}

\section*{Supporting Information}
An R package BayesMetaPenetrance implementing the proposed method
is available at \url{https://personal.utdallas.edu/~sxb125731/} and \url{https://github.com/LakshikaRuberu}.

\end{document}


\maketitle 
\setcounter{page}{20}
\subsection*{Web Appendix A: Assumptions and additional details for the model, simulations, and application}
\subsubsection*{Assumptions  }
\begin{enumerate}[label*=\arabic*.]
    \item Methods - For study $s$ 
 \begin{enumerate}[label*=\arabic*.]

\item Studies reporting RR $y_s^{RR}$ with variance $w_{s}$. 
\begin{itemize}

    \item $q_0(a) = N(A_1, V_1)$: Distribution of age of onset for carriers
    \item $q_1(a) = N(A_0, V_0)$: Distribution of age of onset for non-carriers
\end{itemize}

\item Studies reporting SIR $y_s^{SIR}$ with variance $w_{s}$. 

\begin{itemize}
    \item Assume incidence of BC in the general population is the same as in non-carriers, which is reasonable for rare gene mutations.
\end{itemize}

\item Studies reporting OR $y_s^{OR}$ with variance $w_{s}$.

\begin{itemize}
   
    \item $q_{c0}(a) = N(A_{c0}, V_{c0}) ,q_{c1}(a) = N(A_{c1}, V_{c1})$: Distribution of age of onset for carriers and non-carriers, respectively among cases. 
    \item $q_{h0}(a) = N(A_{h0}, V_{h0}) ,q_{h1}(a) = N(A_{h1}, V_{h1})$: Distribution of age at inclusion in the study for carriers and non-carriers, respectively among healthy controls. 
    \item Usually case-control studies only provide the mean age at diagnosis for cases and the mean age for controls. Therefore, assume $q_{c1}(a) = q_{c0}(a)=q_{c}(a)$ and $q_{h1}(a) = q_{h0}(a)=g_{h}(a)$ leading to
\begin{equation*}
  \log y_s^{OR} \sim N \Biggl(
  \log  \left(  \frac{\int{f_{s}(a|\kappa_s, \lambda_s) q_{c}(a)da}}{\int{f_{0}(a) q_{c}(a)da}} \middle/\frac{\int{(1-F_{s}(a|\kappa_s, \lambda_s)) g_{h}(a)da}}{\int{(1-F_{0}(a))g_{h}(a)da}} \right  ) ,
  w_s^* \Biggr ).
\end{equation*}
\end{itemize}
\end{enumerate}

\item Simulation Study
\begin{enumerate}[label*=\arabic*.]
\item Data Generation
\begin{itemize}
    
    \item In case-control studies, the cases and controls are usually  matched by age. Therefore, for input to the meta-analysis methods, we set mean and SD for the cases same as the corresponding ones for the controls.
\end{itemize}

\end{enumerate}

\item Application

\begin{enumerate}[label*=\arabic*.]

\item If the mean and variance of the age of healthy controls were not reported by the paper, we further assumed that $q_{c}(a) = g_{h}(a)$.
\item Whenever a study did not report any age related summary, $N(63,14.00726)$ is used.

\end{enumerate}
\end{enumerate}

\subsubsection*{Model Details}
For studies reporting a vector of penetrance values $y_s^{\mbox{P}}$,
\begin{itemize}
    \item $W^*$: $m \times m$ covariance matrix of logit($y_s^{\mbox{P}}$)$\\$
    Typically, studies only report 95$\%$ confidence intervals (CIs) of the individual components of $y_s^{\mbox{P}}$. We take the logit of the lower and upper limits of each age-specific CI and estimate variances in the diagonal of $W^*$ considering the interval width and assuming normality.
    
    To compute the off diagonal elements of $W^*$, at each age $m=$ (40, 50, 60, 70, 80), we generate a large number ($n=10,000,000$) of random variables ($\bold{X_m}$) from a normal distribution with mean logit($y_{ms}$) and corresponding variance calculated using 95$\%$ confidence intervals (CIs). Next, for $i^{th}$ draw $(i=1,\dots,n)$ and each age, monotonocity across ages are checked ($X_{i_{40}} < X_{i_{50}} <X_{i_{60}} < X_{i_{70}}< X_{i_{80}}$). We discard draws which do not satisfy this monotonocity. Finally, the covariance of penetrances between any two ages, say $40$ and $50$ are approximated by covariance($X_{40},X_{50}$). 

    \item Note that $m$ can vary across studies even though we use a common $m$ for simplicity of exposition. In fact, these types of studies typically report penetrances at all ages in a wide range (e.g., 35 --- 85) via a curve and thus one can extract estimates at as many ages as desired in that range. The proposed method can accommodate varying $m$ without requiring any change.

\end{itemize}

\subsection*{Web Appendix B: Derivation of equations (1), (2) and (3) used in Section 2}

These equations were given by \citet{Marabelli2016}. Here we present their detailed derivation for the sake of completeness. Let $g=1$ and $g=0$ denote carriers and non-carriers, respectively. \citet{Marabelli2016} assumed that for carriers in study $s$, penetrance of cancer at time $a$ is given by the Weibull cdf $F_1(a|\kappa,\lambda)$ with the associated pdf $f_1(a|\kappa,\lambda)$ where subscript 1 denotes carriers. Note that \citet{Marabelli2016} assume a common Weibull cdf for all studies whereas our proposed method assumes study-specific Weibull cdf denoted by $F_s(a|\kappa_s,\lambda_s)$. The derivation below holds for each study and hence we suppress the subscript $s$ for simplicity and use $F_1(a)$ (subscript 1 for $g=1$) to be consistent with \citet{Marabelli2016}. Further, penetrance at time $a$ for non-carriers is also assumed to follow a Weibull cdf  $F_0(a)$ with associated pdf $f_0(a)$ (The parameters of $F_0(a)$ are assumed to be known). 
$\\$
$\\$
{\bf{\emph{Studies Reporting Relative Risk}}}
\begin{align*}
RR&=\cfrac{P(\text{cancer}|g=1)}{P(\text{cancer}|g=0)}.
\end{align*}
Let the distributions of age for carriers and non-carriers in the study be given by $G(a|g), g=0,1$.
Assume $G(a|g)$ is known. Then,  on average

\[P(\text{cancer}|g) = E_a(P(\text{cancer by age a}|g)).\]

Here $P(\text{cancer by age a}|g)=F_g(a)$, a function of current age $a$. Then

\[E_a(P(\text{cancer by age a}|g))=E_a(F_g(a))=\int F_g(a) G(a|g)da.\]

For $g=1$,
\[P(\text{cancer}|g=1)=\int F_1(a) G(a|g=1)da \text{ and}\] 

For $g=0$,
\[P(\text{cancer}|g=0)=\int F_0(a) G(a|g=0)da.\]
$\\$
Therefore,
\[RR=\cfrac{P(\text{cancer}|g=1)}{P(\text{cancer}|g=0)}=\cfrac{\int F_1(a) G(a|g=1)da}{\int F_0(a) G(a|g=0)da}.\]

$\\$
However, papers usually do not report $G(a|g)$. Therefore, we make use of the age of onset distribution for carriers and non carriers denoted by $q(a|g), g=0,1$. We approximate the cancer rate as 
\[P(\text{cancer}|g) \approx P(\text{cancer onset at any age}|g)=E_a(P(\text{cancer onset at age a}|g)).\]

Here $P(\text{cancer onset at age a}|g)=f_g(a)$, a function of age of onset $a$. Then
\[E_a(P(\text{cancer at age a}|g))=E_a(P(\text{cancer|age of onset=a, } g ))=E_a(f_g(a))=\int f_g(a) q(a|g)da.\]

For study $s$
\[RR \approx \cfrac{\int f_1(a) q(a|g=1)da}{\int f_0(a) q(a|g=0)da}.\]
Let $q(a|g=1)=q_1(a)$ and $q(a|g=0)=q_0(a)$. Then 
\begin{equation}
\label{RR}
    RR \approx \cfrac{\int f_1(a) q_1(a)da}{\int f_0(a) q_0(a)da}.
\end{equation}
$\\$
{\bf{\emph{Studies Reporting Standard Incidence Ratio}}}$\\$
\[SIR=\cfrac{P(\text{cancer}|g=1)}{P(cancer)}.\]
By the law of total probability 
\[P(\text{cancer})=P(g=1)\times P(\text{cancer}|g=1)+P(g=0) \times P(\text{cancer}|g=0). \]
$\\$
Now by equation (\ref{RR})
\begin{equation}
\label{SIR}
SIR=\cfrac{\int f_1(a) q_1(a)}{P(g=1)\times \int f_1(a) q_1(a)+P(g=0) \times \int f_0(a) q_0(a)}.
\end{equation}

However, $P(g=1)$ may not be well established and for a rare pathogenic variant $\\P(g=1)<<P(g=0)$. Therefore we can approximate 
\[P(\text{cancer})\approx P(\text{cancer}|g=0).\]
With this approximation the expression for SIR reduces to that of RR in \ref{RR}.

$\\$
{\bf{\emph{Studies Reporting Odds Ratio}}}
$\\$
Let
\begin{align*}
P_{g|c}&=P(g=1|\text{cancer  at any age})\\
P_{g|h}&=P(g=1|\text{healthy at age of inclusion in study})
\end{align*}
 \[OR=\cfrac{ \cfrac{P_{g|c}}{1- P_{g|c}}}{ \cfrac{P_{g|h}}{1- P_{g|h}}}.\]

$\\$
By Bayes' Theorem

\begin{align*}
    P_{g|c}&=\cfrac{P(g=1)P(\text{cancer at any age a}|g=1)}{\sum_{j=0}^1 P(g=j)P(\text{cancer at any age a }|g=j)}\\
    \cfrac{P_{g|c}}{1- P_{g|c}}&=\cfrac{P(g=1)P(\text{cancer at any age a}|g=1)}{P(g=0)P(\text{cancer at any age a}|g=0)}
\end{align*}
$\\$
Following the same steps as used in deriving equation (\ref{RR}), 
\begin{subequations}
\label{eq:subeqns}
\begin{align}
    \cfrac{P_{g|c}}{1- P_{g|c}}=\cfrac{P(g=1)\int f_1(a) q_{c1}(a)da}{P(g=0)\int f_0(a) q_{c0}(a)da} ,\label{eq:subeqns1}
\end{align}
where $q_{c1}(a)$ and $q_{c0}(a)$ are distributions of age of onset of cancer for carriers and non-carriers respectively. Similarly,
\begin{align*}
    P_{g|h}&=\cfrac{P(g=1)P(\text{healthy at age of inclusion in study}|g=1)}{\sum_{j=0}^1 P(g=j)P(\text{healthy at age of inclusion in study}|g=j)} \; \; \text{ and}\\
    \cfrac{P_{g|h}}{1- P_{g|h}}&=\cfrac{P(g=1)P(\text{healthy at age of inclusion in study}|g=1)}{P(g=0)P(\text{healthy at age of inclusion in study}|g=0)}.
\end{align*}
Let $q_{h1}(a)$ and $q_{h0}$ be distributions of age at inclusion in the study for carriers and non-carriers, respectively. Following the derivation of equation~ (\ref{RR}),
\begin{align*}
P(\text{healthy at age of inclusion in study}|g) = \int (1-F_g(a))q_{hg}(a) da,
\end{align*} 
where $1-F_g(a),\; g=0,1$ is the disease-free survival function by age $a$. Then for study $s$,
\begin{align}
 \cfrac{P_{g|h}}{1- P_{g|h}}&=\cfrac{P(g=1)\int(1-F_{1}(a)) q_{h1}(a)da}{P(g=0)\int(1-F_{0}(a))q_{h0}(a)da}.\label{eq:subeqns2}
\end{align}
Now by (\ref{eq:subeqns1}) and (\ref{eq:subeqns2}),
\begin{equation}
\label{OR}
     OR=\left. \frac{\int{f_{1}(a) q_{c1}(a)da}}{\int{f_{0}(a) q_{c0}(a)da}} \middle/\frac{\int{(1-F_{1}(a)) q_{h1}(a)da}}{\int{(1-F_{0}(a))q_{h0}(a)da}}. \right.
\end{equation}

\end{subequations}
\subsection*{Web Appendix C: Details on determining limits of the hyper-prior distributions}

Our hierarchical prior structure is as follows: $\pi(\kappa_s|a, b) =$ Gamma$(a, b), \;  \pi(\lambda_s|c, d) =$ Gamma$(c, d),\;$ where $a$ and $c$ are shape parameters and $b$ and $d$ are scale parameters and $\pi(a|l_a, u_a) =$ U$(l_a, u_a), \\ \pi(b|l_b, u_b) =$ U$(l_b, u_b), \;  \pi(c|l_c, u_c) =$ U$(l_c, u_c), \;$and$ \; \pi(d|l_d, u_d) =$ U$(l_d, u_d) $. The lower and upper limits $(l, u)$ of these uniform distributions are pre-specified. In our simulations and application, we use $a \sim$ U$(7.5,27.5)$ , $b \sim$ U$(0.15,0.25)$, $c \sim$ U$(43,63)$, and $d \sim$ U$(1.32,2.02)$. We determine these limits based on the consideration that the Weibull penetrance distributions arising from these choices cover a wide possibility of penetrance values at ages 20 - 80 while ensuring that unrealistically large values are unlikely. To this end, we first fix $(l, u)$ values for each hyper-parameter and generate 50 values of $a$, $b$, $c$, and $d$ from their respective uniform priors. Then we generate 50 $\kappa_s$ and 50 $\lambda_s$ values from Gamma$(a, b)$ and Gamma$(c, d)$ priors and compute penetrance values from  the resulting 2500 Weibull$(\kappa_s,\lambda_s)$ distributions. Finally, we draw histograms of penetrance values at ages 20, 40, 60, and 80. We repeat this process by varying the limits $(l, u)$ for each of the four hyper-parameters and choose the ones that best satisfy our consideration. This process resulted in the uniform distributions with bounds as specified above. The histograms of penetrance corresponding to the final chosen limits are shown in Web Figure 3. Note that the penetrance values and their frequencies as reflected in these histograms cover a wide range matching what we expect for most cancer genes. Hence, the above prior specifications should be portable to most gene-cancer applications.

\subsection*{Web Appendix D: Details of the MCMC algorithm}
At iteration $(t+1)$ of the algorithm, the following updates are made.$\\$
{\bf {Update of $\mathbf{\kappa_s}$:}} The full conditional distribution of $\kappa_s$ is given by 
\[ \pi(\kappa_s|\by_s, \lambda_s, a, b) \propto L(\btheta_s)\pi(\kappa_s|a, b). \]
We use a Gamma$(\alpha_s, \beta_s)$ proposal whose mean $\alpha_s\beta_s$ is set to be equal to $\kappa_s^{(t)}$ (the current value of $\kappa_s$) and variance $\alpha_s\beta_s^2$ is set to be equal to a function of the current value of $\kappa_s^{(t)}$. In particular, for studies reporting age-specific penetrance, $\alpha_s\beta_s^2 =$ log$_{500}(\kappa_s^{(t))}+0.01)$, for studies reporting RR, it is log$_{500}(\kappa_s^{(t))}+2000)$ and for those reporting OR, it is log$_{500}(\kappa_s^{(t))}+200000)$. These choices ensure convergence and recommended acceptance rates \cite[Chapters~11-12]{gelman1995bayesian}.$\\$
$\\$
{\bf {Update of $\mathbf{\lambda_s}$:}} $\lambda_s$ is updated in a similar way as $\kappa_s$
using a Gamma proposal whose mean is $ \lambda_s^{(t)}$ and variance is set to be equal to $\lambda_s^{(t)^{0.4}}, \lambda_s^{(t)^{0.9}},$ and $\lambda_s^{(t)^{1.2}}$ for studies reporting age-specific penetrance, RR, and OR, respectively.  $\\$
$\\$
{\bf {Update of $a$:}} The conditional distribution of $a$ is given by 
\begin{align*}
\pi(a|\by_s, \bkappa, b) & \propto \pi(\bkappa|a, b)\pi(a) \\
& \propto \left(\frac{1}{\Gamma(a) b^{a}}\right)^S \prod_{s=1}^{S} (\kappa_s)^{a-1}\exp{\left(-\sum_{s=1}^{S}\kappa_s/b\right)} I(l_a < a < u_a) \\
& \propto  \left(\frac{1}{\Gamma(a) b^{a}}\right)^S \prod_{s=1}^{S} (\kappa_s)^{a-1} I(l_a < a < u_a). 
\end{align*}
We use a uniform proposal distribution U$(\max(l_a, a^{(t)} - 9), \; \min(a^{(t)} + 9, u_a))$.$\\$
$\\$
{\bf {Update of $b$:}} The conditional distribution of $b$ is given by 
\begin{align*}
\pi(b|\by_s, \bkappa, a) & \propto \pi(\bkappa|a, b)\pi(b) \\
& \propto \left(\frac{1}{\Gamma(a) b^{a}}\right)^S \prod_{s=1}^{S} (\kappa_s)^{a-1}\exp{\left(-\sum_{s=1}^{S}\kappa_s/b\right)} I(l_b < b < u_b) \\
& \propto  \left(\frac{1}{b^{a}}\right)^S  \exp{\left(-\sum_{s=1}^{S}\kappa_s/b\right)} I(l_b < b < u_b). 
\end{align*}
We use a uniform proposal distribution U$(\max(l_b, b^{(t)} - 0.04), \; \min(b^{(t)} + 0.04, u_b))$.$\\$
$\\$
{\bf {Update of $c$ and $d$:}} These are carried out in a similar manner as for $a$ and $b$ with uniform proposal distributions of U$(\max(l_c, c^{(t)} - 8), \; \min(c^{(t)} + 8, u_c))$ and $\\$U$(\max(l_d, d^{(t)} - 0.22), \; \min(d^{(t)} + 0.22, u_d))$, respectively.

The convergence of the algorithm is assessed using Gelman-Rubin statistic and trace plots \cite[Chapters~11-12]{gelman1995bayesian}. Web Figure 4 provides trace plots and the corresponding Gelman-Rubin statistic for all the parameters in a typical simulation replicate.

\subsection*{Web Appendix E: Method by \citet{Marabelli2016}}
\label{sub:Marab}

Marabelli et al. express the OR, RR, and SIR in terms of penetrance parameters $\kappa$ and $\lambda$, assumed to be the same across all studies, leading to a fixed-affects approach. Risk measures (penetrance or their function such as OR) are assumed to be normally distributed with mean and SD fixed at the estimate and SD reported by the study. For example, for a study that reports OR, it is assumed that OR, which is a r.v given by equation~\eqref{OR}, follows a normal distribution with mean and SD pre-specified and obtained based on estimate of OR and its CI reported by the study. Studies are assumed to be independent. The overall normal likelihood is then maximized with respect to $\kappa$ and $\lambda$ to obtain the maximum likelihood estimates (MLE). The Weibull distribution with these MLEs of $\kappa$ and $\lambda$ is the meta-analysis curve. 95$\%$ CrIs are also obtained using a numerical method.

\subsection*{Web Appendix F: Details about data generation model}
\subsubsection*{Simulations  based on ATM}

As mentioned in Section 3.1.1, we generate study-specific Weibull parameters from normal distributions $\kappa_s \sim N(4.55,0.525)$ and $\lambda_s \sim N(95.25,12.375)$. To obtain these normal distribution parameters, we first approximate penetrance curve for each of the 17 studies by a Weibull distribution. For the studies reporting age-specific penetrance this is trivial. For studies, reporting RR, SIR, or OR we assume that the reported measure remains constant over patient's life-time and then combine it with the age-specific penetrance for non-carriers obtained from SEER \citep{seer} to get age-specific penetrance curves using the following formulas \citep{ASK2ME}. We define
\begin{align*}
T&=\text{Age of first cancer}\\
g&=\text{Carrier status; g = 1 for a carrier of pathogenic variant, g = 0 for a non-carrier}
\end{align*}
The absolute risk of cancer for gene mutation carriers between age $t_1$ and $t_2$ is given by
\[P(t_1\leq T<t_2| g=1).\]
The formula for calculating age-specific penetrance values for studies reporting RR and SIR is given by
\begin{align*}
RR&=\frac{P(t_1\leq T<t_2| g=1)}{P(t_1\leq T<t_2| g=0)} \\
\implies P(t_1\leq T<t_2| g=1)&=RR\times {P(t_1\leq T<t_2| g=0).} 
\end{align*}
\noindent
For studies reporting OR, the formula is
\begin{align*}
OR&=\cfrac{\cfrac{P(t_1\leq T<t_2| g=1)}{1-P(t_1\leq T<t_2| g=1)}}{\cfrac{P(t_1\leq T<t_2| g=0)}{1-P(t_1\leq T<t_2| g=0)}}\\
\implies P(t_1\leq T<t_2| g=1)&=\cfrac{OR\times\cfrac{P(t_1\leq T<t_2| g=0)}{1-P(t_1\leq T<t_2| g=0)}}{1+OR\times \cfrac{P(t_1\leq T<t_2| g=0)}{1-P(t_1\leq T<t_2| g=0)}},
\end{align*}
where $P(t_1\leq T<t_2| g=0)$ is obtained from the age-specific risks reported by SEER. 

The resulting age-specific penetrance curves are then approximated by Weibull distributions. Thus, we get 17 pairs of ($\kappa,\lambda$) values, one per study. We use the mid-point and the range of these ($\kappa,\lambda$) values to calculate the parameters of the respective normal distributions. More specifically, we get a range of $\kappa$ values from 3.5 to 5.6. Thus, the mid-point is 4.55 and assuming that the range covers $\pm2SD$ from the mid-point, the SD is 0.525, which resulted in $\kappa_s \sim N(4.55,0.525)$. Similarly, the range of $\lambda$ values is 70.5 to 120, resulting in $\lambda_s \sim N(95.25,12.375)$.

Note that the data generation model described above is different from the proposed (i.e., working) model even though both assume that the study-specific penetrance follows a Weibull distribution (assumed by \citet{Marabelli2016} as well). First, the likelihood in the working model assumes normal distributions for risk estimates while in data generation, risk estimates are not generated directly from any distribution. Next, in the data generation model,  $\kappa_s$ and $\lambda_s$ parameters are generated from normal distributions while our hierarchical working model assumes that $\kappa_s$ and $\lambda_s$ follow gamma distributions.

\subsubsection*{Simulations  based on PALB2}
For the simulations based on PALB2, we follow the same procedure as described above. We approximate the estimates from the four real studies by Weibull penetrance curves. The range of $\kappa$ values is 3 --- 4.4 and that of $\lambda$ values is 70 --- 99, which result in the normal distributions of N$(3.7, 0.35)$ for $\kappa_s$ and N$(84.5, 7.25)$ for $\lambda_s$.

\subsection*{Web Appendix G: Details and Results of Sensitivity Analysis}
\label{sub:sens}
To determine how sensitive our results are to the choice of various fixed quantities, we conduct various sensitivity analyses. First, we investigate how sensitive our results are to the choice of fixed hyper-parameters. We explore two other sets of hyper-priors denoted by Set 1 and Set 2. Recall that  our hyper-priors for $a,b,c,$ and $d$ parameters are uniform distributions and the specific choices of lower and upper limits result in realistic distributions of penetrance values at different ages (as described in Appendix B). To study sensitivity, we vary these limits in such a way that Set 1 would allow for a wider range of penetrance values at each age while Set 2 would somewhat restrict the range of penetrance values. Both sets still ensure that the resulting penetrance values are in realistic ranges. The two sets of hyper-priors are Set 1: $a\sim $U$(6.5,26.5)$, $b\sim $U$(0.14,0.24)$, $c\sim $U$(38,58)$,  $d\sim $U$(1.39,1.99)$ and Set 2: $a\sim $U$(10.5,30.5)$, $b\sim $U$(0.18,0.28)$, $c\sim $U$(49,69)$,  $d\sim $U$(1.31,1.91)$. Web Figure 5 shows the histograms of penetrance resulting from these two sets. The results of the sensitivity analysis are shown in Web Table 3. As compared to Table 1 in manuscript, there are some differences in penetrance estimates, MSEs and coverage of $95\%$ credible intervals, however, they do not differ substantially. Thus, we may conclude that the results from the proposed method are relatively robust to the choice of priors as long as they capture a realistic range of penetrance values.

Next, we examine the sensitivity of our final estimates to different age-related distributions needed in equations \ref{RR}, and \ref{OR} i.e., $q_1$, $q_0$, $q_{c}$, and, $g_{h}$. Note that the simulation results under Scenarios 1 and 2 present some evidence of robustness. Recall that in Scenario 1, the mean and SDs of the age distributions vary across the RR/OR studies assumed to report age-related summaries and it is set to $N(63, 14.00726)$ for other RR/OR studies. While in Scenario 2, age distributions for all RR/OR studies are set to be $N(63, 14.00726)$. This also shows that our method is not sensitive to non-reporting of age-related summaries by studies under consideration. When such information is available, it can be (and should be) included. Otherwise readily available summaries from population-based registries like SEER can be plugged in. To further study the robustness, we conduct a simulation where we assume all studies report the relevant age related summaries (note these are needed only for studies reporting RR, SIR, and OR). That is, none of the distributions are fixed at $N(63, 14.00726)$. The mean (SD) of $q_1$, $q_0$, and $q_c$ (we set $q_c$ = $g_h$ in our simulations) generated across 500 simulations replicates were 67.75 (4.87), 67.60 (1.82), and 67.48 (1.67), respectively. The results (shown in Web Table 4) are similar to our earlier results for Scenarios 1 and 2 indicating that our method is robust to different age distributions.

Finally, we investigate the robustness of our method to varying values of $m$, the number of time points at which penetrances are reported by different studies. Note that the two ATM-BC studies that reported age-specific penetrance (listed in Table 1), in fact, presented the full penetrance curve over a wide range and so penetrance at any number of ages can be extracted. In our simulations (and actual meta-analysis), we used penetrances at ages 40 --- 80 with 10-year interval for both studies, i.e., $m$ = 5. To study sensitivity of our results to this choice, we let one  study have $m = 6$ at ages 32, 42, 52, 62, 72, and 82 while $m$ = 5 for the other study as before. The results are given in Web Table 5 --- they are practically the same as the ones given in Table 1 and Web Table 1.

\subsection*{Web Appendix H: Assumptions Specific to Analysis of the ATM Gene}
Following \citet{Marabelli2016} and our simulations, for studies reporting SIR, we use equation ~\ref{RR} instead of ~\ref{SIR}. Furthermore, we assume $q_{c1}(a) = q_{c0}(a)=q_{c}(a)$ and $q_{h1}(a) = q_{h0}(a)=g_{h}(a)$ (as defined after equation ~\eqref{OR}). Moreover, if the mean and variance of the age of healthy controls were not reported by the paper, we further assumed that $g_{h}(a) = q_{c}(a)$. Whenever a study did not report any age related summary we use a mean age of 63 and SD 14.00726 as used in our simulations. For case-control studies in which no mutations were reported in controls (i.e., OR is not defined), we add $0.5$ to each cell of the $(2 \times 2)$ table to estimate OR and its SE \citep{jb1956estimation,gart1967bias}.
In a Bayesian meta-analysis based on individual-level data this step would not be necessary, but here we are modeling the situation in which only summary data is available from the literature.

 \bibliography{arxiv_supplementary}

\begin{table} [h]
\centering
\caption{Simulation Results for Setting 1 (ATM-based) and Scenario 2}
\begin{tabular}{lccccc}\hline
Age & 40 & 50 & 60 & 70 & 80\\ \hline
True Penetrance& 0.026 &	0.067&	0.141&	0.253&	0.398\\
Estimated Penetrance (Bayesian) &0.027&	0.065&	0.136&	0.247&	0.398\\
Estimated Penetrance (Marabelli) & 0.040&	0.081&	0.141&	0.222&	0.324\\
MSE (Bayesian) & 0.0001	&0.0004	&0.0009	&0.0020	&0.0036\\
MSE (Marabelli) & 0.0017 &	0.0036 &	0.0054 &	0.0076	 &0.0124 \\
95\% CrI coverage (Bayesian) &0.980&	0.980&	0.970&	0.952&	0.938\\ 
95\% CrI coverage (Marabelli) &0.162&	0.140&	0.104&	0.082&	0.050\\\hline
\end{tabular}
\end{table}

\begin{table}[h]
\centering
\caption{Simulation Results for Setting 2 (PALB2-based) and Scenario 2}
\begin{tabular}{lccccc}\hline
Age & 40 & 50 & 60 & 70 & 80\\ \hline
True Penetrance & 0.066 & 0.143 & 0.257 & 0.404 & 0.565\\
Estimated Penetrance (Bayesian) & 0.071	&0.148&	0.263&	0.411	&0.574\\
Estimated Penetrance (Marabelli) & 0.070&	0.154&	0.269&	0.403&	0.544\\
MSE (Bayesian) & 0.0002&0.0004&	0.0009&	0.0015&	0.0019\\
MSE (Marabelli) & 0.0004&	0.0011	&0.0018	&0.0022&	0.0029 \\
95\% CrI coverage (Bayesian) &1.00& 	1.00& 	1.00	& 1.00	& 0.996\\ 
95\% CrI coverage (Marabelli) & 0.238&	0.190&	0.132	&0.096&	0.152\\\hline
\end{tabular}
\end{table}

\begin{table}
 \centering
\caption{Simulation Results of Sensitivity Analysis Based on Two Different Sets of Hyper-priors -- Set 1 (top) and Set 2 (bottom)}
\label{sens}
\begin{tabular}{lccccc}
  \hline
Age & 40 & 50 & 60 & 70 & 80\\ \hline
True Penetrance& 0.026 &	0.067&	0.141&	0.253&	0.398\\\hline
Estimated Penetrance   & 0.031	&0.074&	0.148&	0.262	&0.412\\
MSE   & 0.0002	&0.0006&	0.0015	&0.0030&	0.0050\\
95\% CrI coverage   &0.990&	0.986&	0.964	&0.950&	0.920\\ \hline
Estimated Penetrance   & 0.024	  &0.060	  &0.127  &	0.235	  &0.387\\
MSE   & 0.0001  &	0.0005  &	0.0013  &	0.0025  &	0.0039\\
95\% CrI coverage &0.958  &	0.954  &	0.952	  &0.942  &	0.928\\ \hline
\end{tabular}
\end{table}

\begin{table}[htbp!]
\centering
\caption{Simulation Results of Sensitivity Analysis for Different Age Distributions - Setting 1 (ATM-based) and Scenario 1}
\begin{tabular}{lccccc}\hline
Age & 40 & 50 & 60 & 70 & 80\\ \hline
True Penetrance& 0.026 &	0.067&	0.141&	0.253&	0.398\\
Estimated Penetrance (Bayesian) &0.028	&0.068	&0.140&	0.251&	0.402\\
MSE (Bayesian) & 0.0002	&0.0006&	0.0015&	0.0029&	0.0049\\
95\% CrI coverage (Bayesian) &0.974&	0.972&	0.964&	0.952&	0.918\\  \hline

\end{tabular}
\end{table}  

\begin{table}[htbp!]
\centering
\caption{Simulation Results of Sensitivity Analysis with the number of time points ($m$) assumed to be 5 and 6 for the two studies reporting age-specific penetrance - Setting 1 (ATM-based) and Scenario 1}
\begin{tabular}{lccccc}\hline
Age & 40 & 50 & 60 & 70 & 80\\ \hline
True Penetrance& 0.026 &	0.067&	0.141&	0.253&	0.398\\
Estimated Penetrance (Bayesian) &0.031&	0.072&	0.145&	0.258&	0.408\\
MSE (Bayesian) & 0.0003	&0.0008	&0.0017	&0.0031&	0.0050\\
95\% CrI coverage (Bayesian) &0.996	&0.990&	0.982&	0.956&	0.934\\  \hline

\end{tabular}
\end{table}

\newpage

\begin{figure}[htp]
    \centering
    \includegraphics[width=\textwidth]{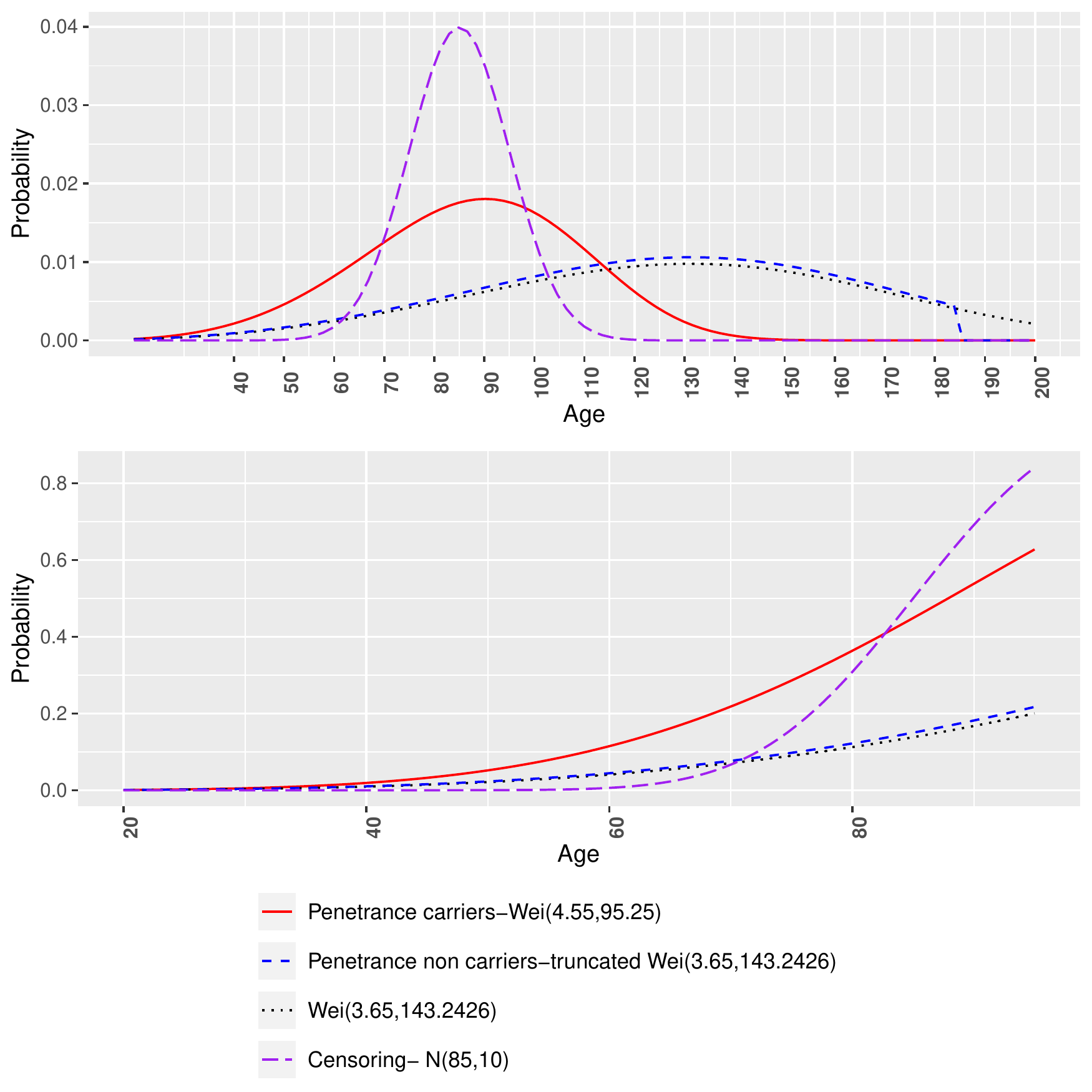}
    \caption{ Weibull and normal distributions for generating carriers and non-carriers in ATM-BC analysis. (Top: pdf and Bottom: cdf)}
\end{figure}

\begin{figure}[h]
    \centering
    \includegraphics[width=\textwidth]{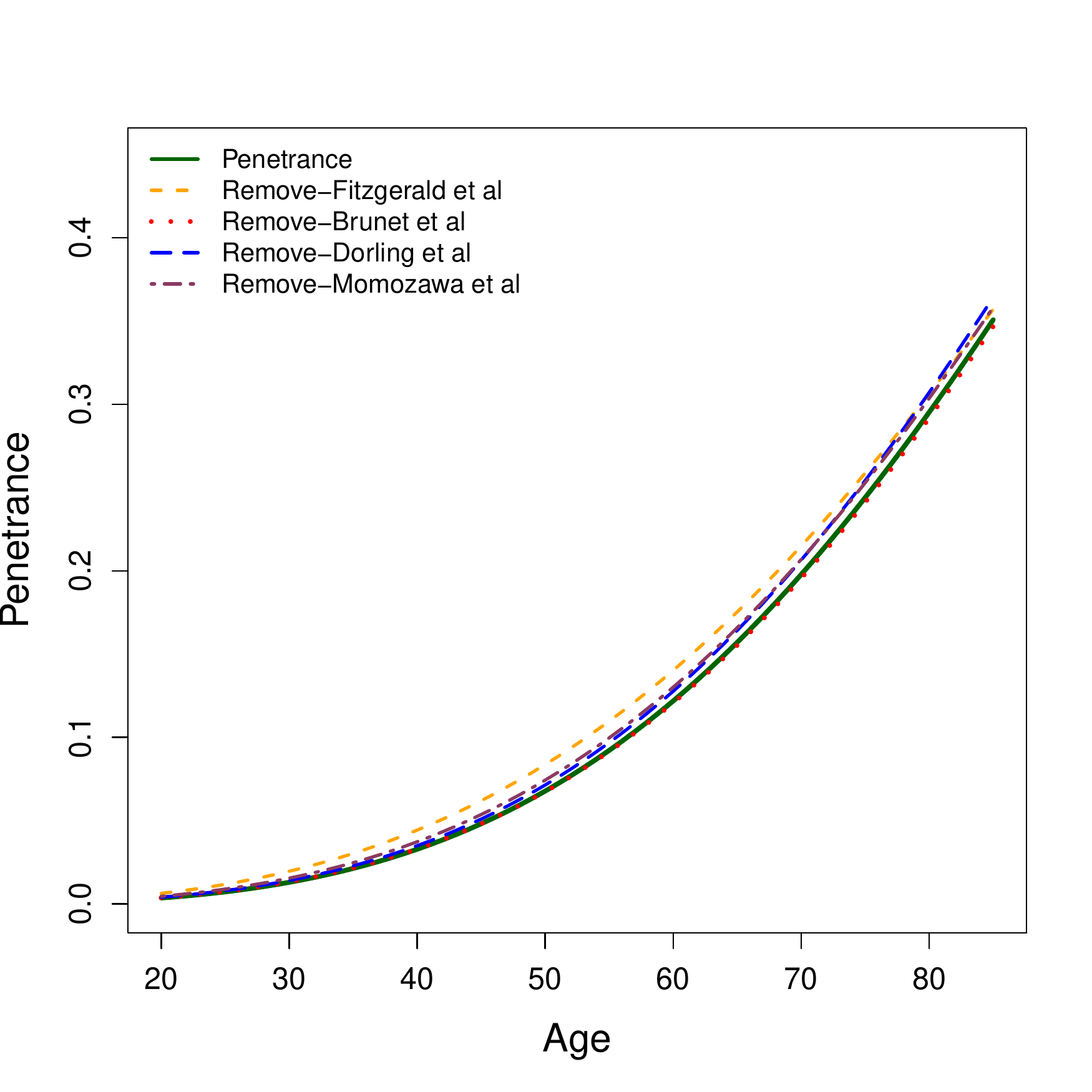}
    \caption{Results of Sensitivity analysis: Removing studies with early onset cases or controls restricted to older ages one at a time}
\end{figure}

\begin{figure}[htbp!]
    \centering
    \includegraphics[width=0.6\textwidth]{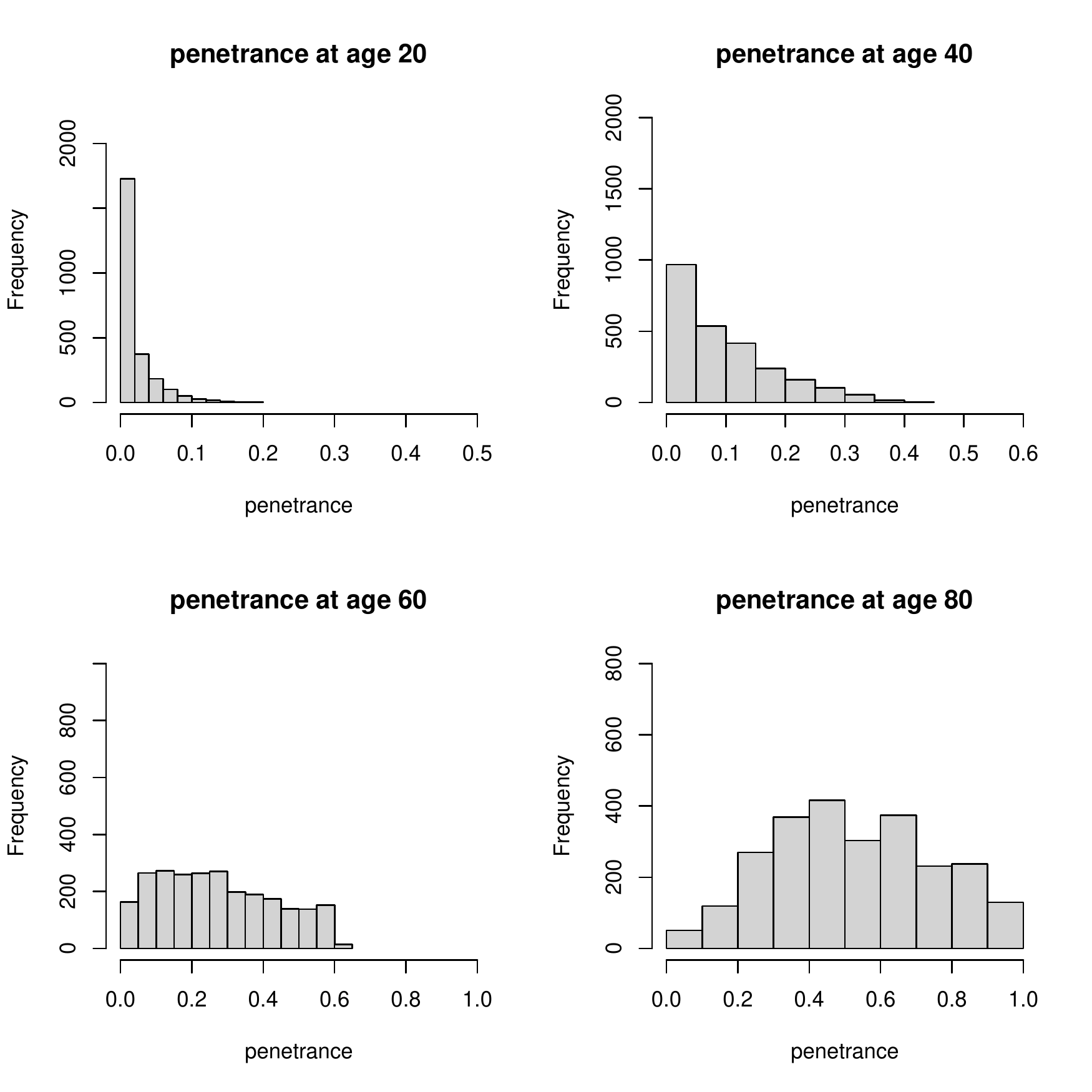}
    \caption{Distributions of penetrance resulting from the specified hierarchical prior distributions}
\end{figure}

\begin{figure}
\centering
\includegraphics[page=1]{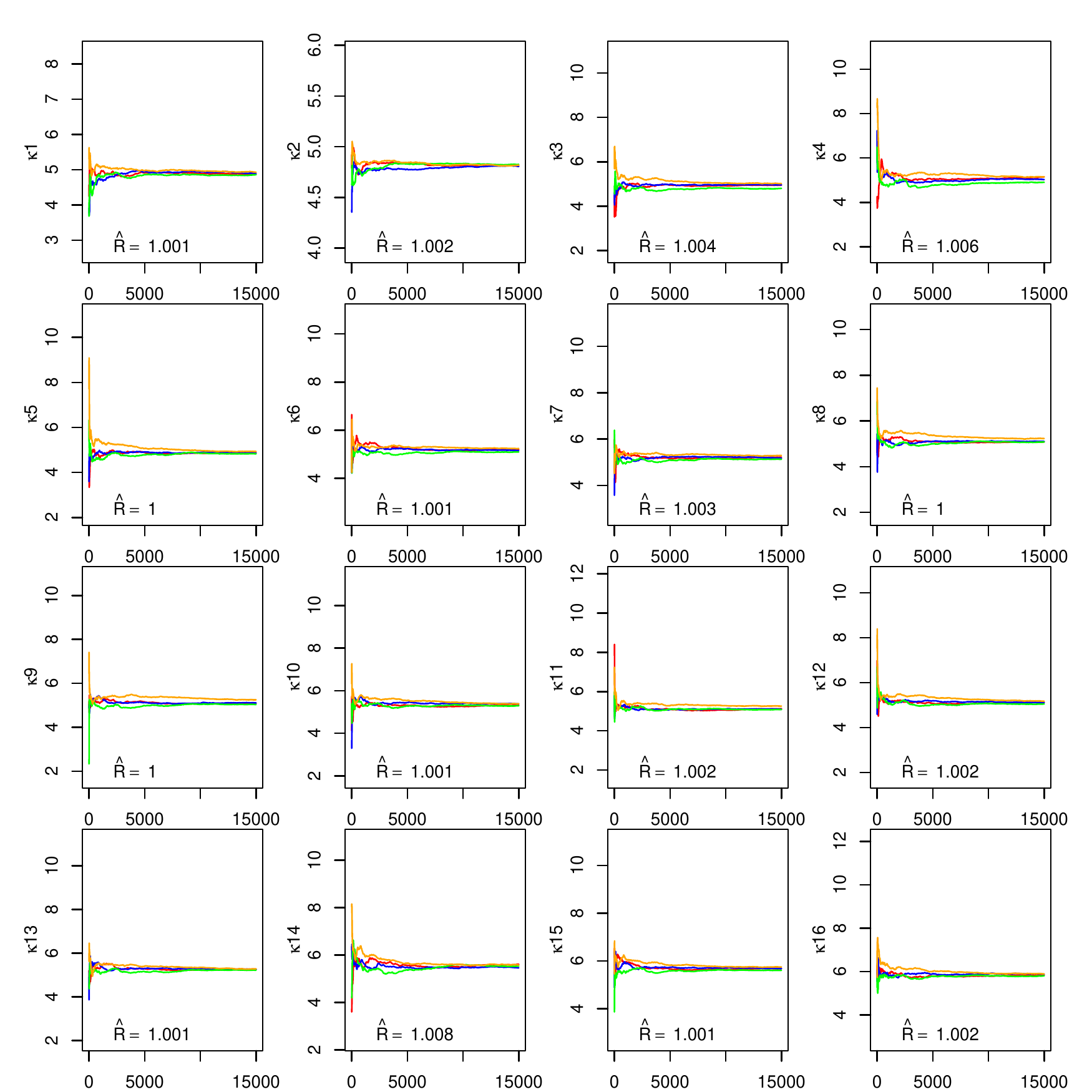}
\caption{Trace plots (part 1) from different starting points and corresponding Gelman-Rubin statistics (indicated by $\hat{R}$ at the bottom of each plot) for a typical sample used in simulations --- Setting 1 (ATM-based) and Scenario 1 }
\end{figure}
\begin{figure}
\ContinuedFloat
\centering
\includegraphics[page=2]{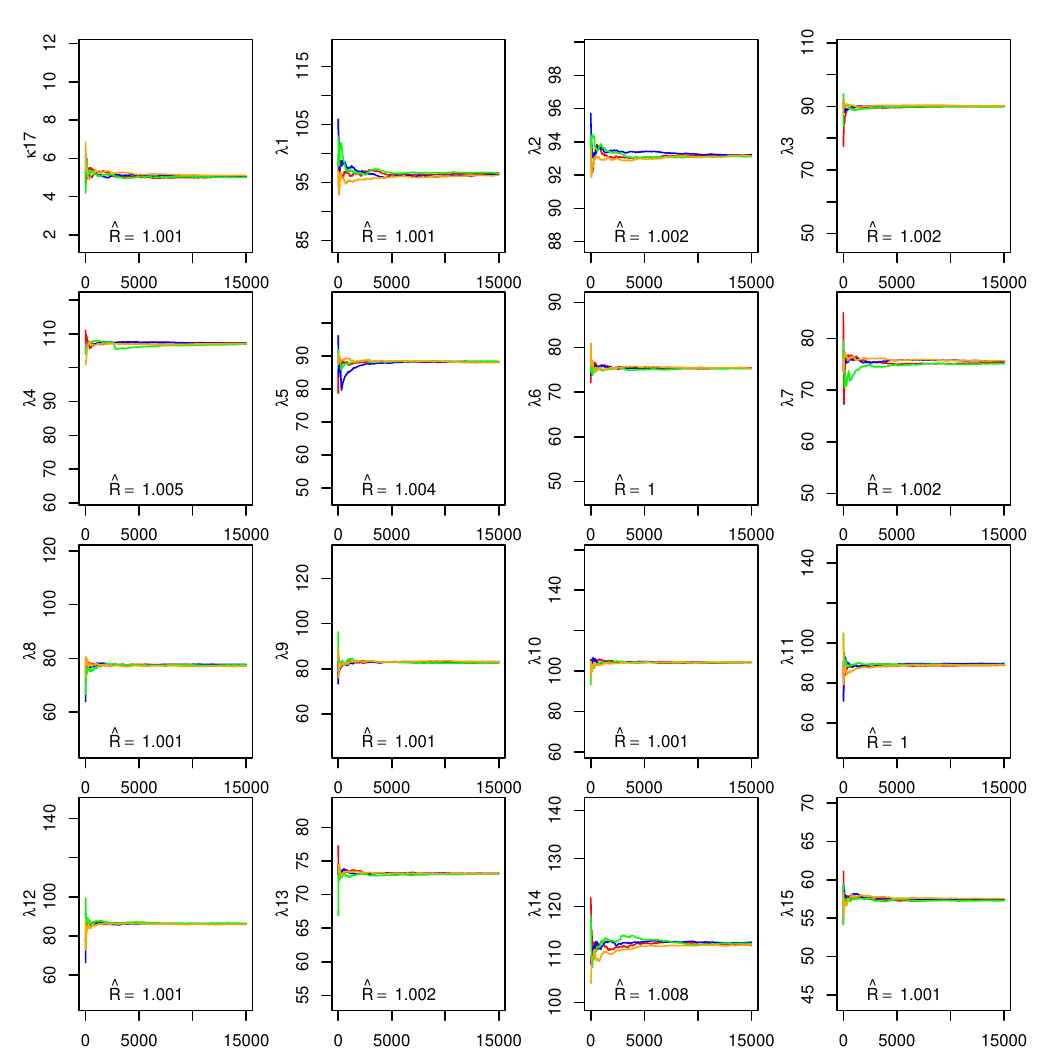}
\caption{Trace plots (part 2) from different starting points and corresponding Gelman-Rubin statistics (indicated by $\hat{R}$ at the bottom of each plot) for a typical sample used in simulations --- Setting 1 (ATM-based) and Scenario 1}
\end{figure}
\begin{figure}
\ContinuedFloat
\centering
\includegraphics[page=3]{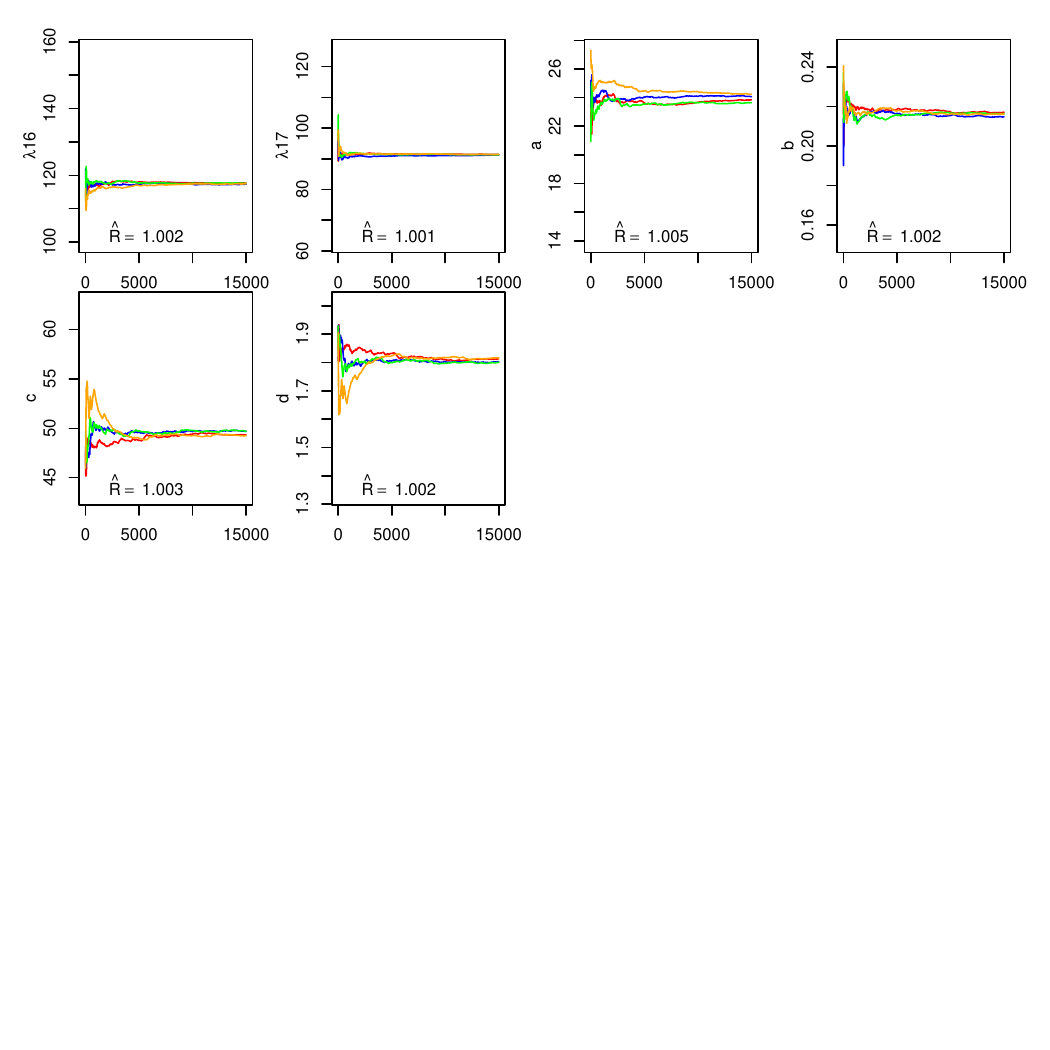}
\caption{Trace plots (part 3) from different starting points and corresponding Gelman-Rubin statistics (indicated by $\hat{R}$ at the bottom of each plot) for a typical sample used in simulations --- Setting 1 (ATM-based) and Scenario 1}
\end{figure}

\begin{figure}[htp]
    \centering
    \subfloat[]{{\includegraphics[width=0.6\textwidth]{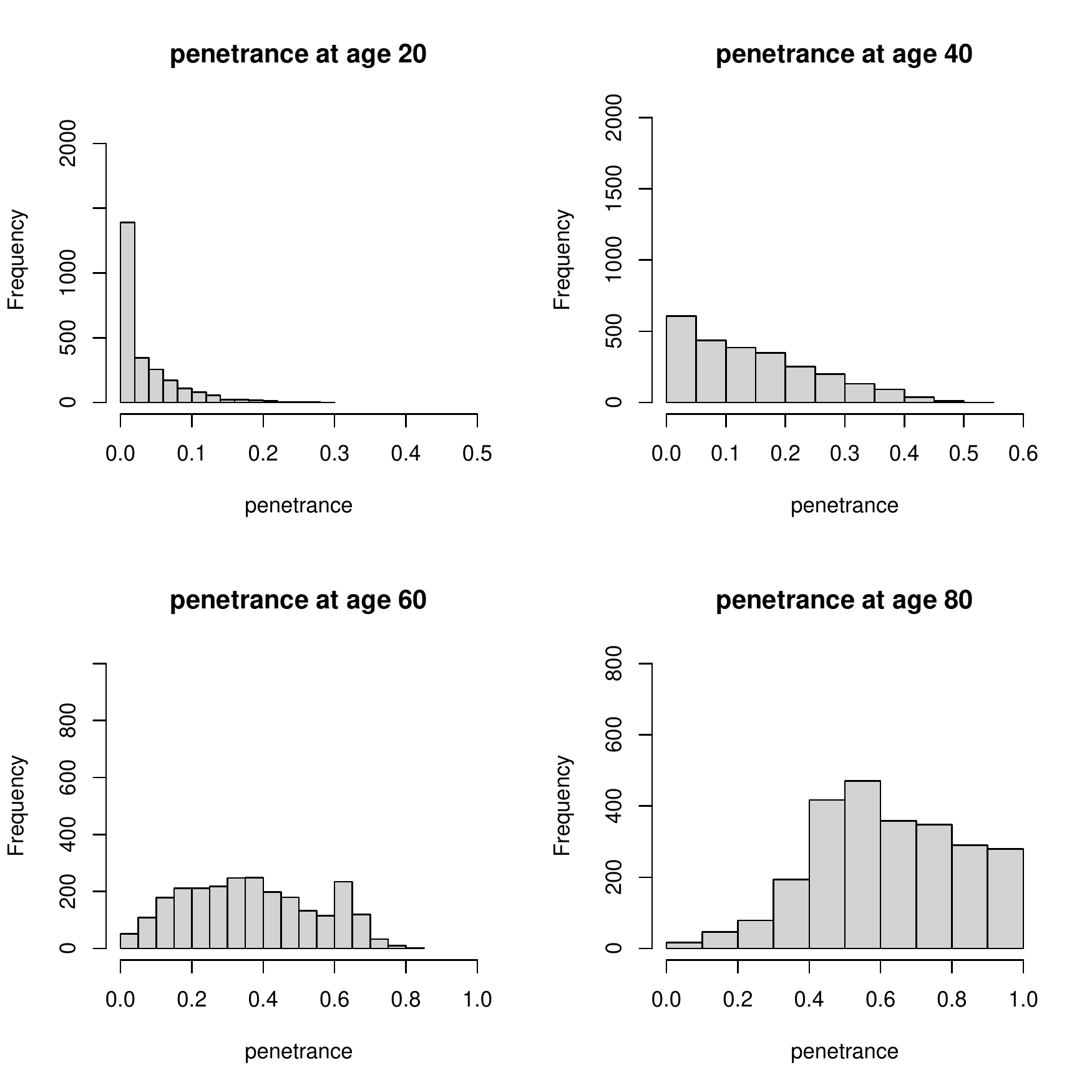} }}
    \qquad
    \subfloat[]{{\includegraphics[width=0.6\textwidth]{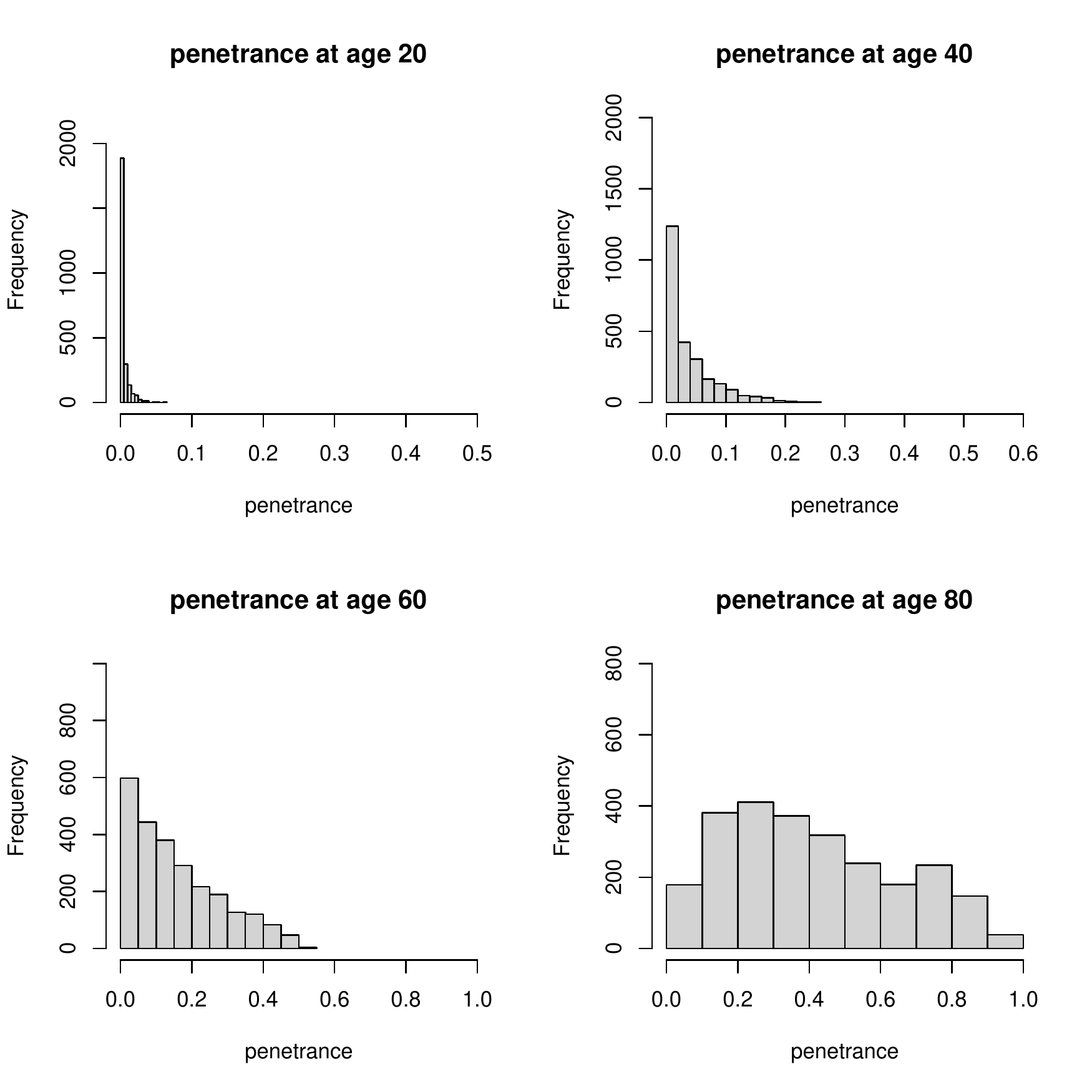} }}
    \caption{Distributions resulting from hierarchical priors used in sensitivity analysis based on (a) Set 1 (b) Set 2 }
    \label{fig:example}
\end{figure}
